\documentclass[12pt]{article}

\usepackage[english]{babel}
\usepackage{natbib}

\usepackage{setspace}
\usepackage{enumitem}
\usepackage{subfig}

\usepackage{amsmath}
\usepackage{amssymb}
\usepackage{amsfonts}
\usepackage{graphicx}
\usepackage{bm}

\usepackage[colorlinks=true, allcolors=blue]{hyperref}

\usepackage{dsfont}
\usepackage{enumitem}
\usepackage{booktabs}
\usepackage{bm}
\usepackage{tikz}
\usepackage[normalem]{ulem}

\usepackage{amsthm}

\newtheorem{theorem}{Theorem}
\newtheorem{proposition}{Proposition}
\newtheorem{lemma}{Lemma}
\newtheorem{corollary}{Corollary}

\theoremstyle{definition}

\newtheorem*{definition}{Definition}  

\newtheorem{observation}{Observation}

\newcommand{\R}{{\mathcal{R}}} 
\newcommand{\M}{{\mathcal{M}}}
\newcommand{\tilP}{{\widetilde{P}}} 
\newcommand{\tilR}{{\widetilde{R}}} 
\newcommand{\tilmu}{{\widetilde{\mu}}}
\newcommand{\hatP}{{\widehat{P}}} 
\newcommand{\hatR}{{\widehat{R}}} 

\newcommand{\st}{{s}} 

\DeclareMathOperator{\SD}{SD}   
\DeclareMathOperator{\RSD}{RSD} 

\newcommand{\circled}[1]{\tikz[baseline=(char.center)]{ 
		\node[shape=circle, draw, inner sep=0pt, minimum size = 6mm](char) {#1};}}

\newcommand{\squared}[1]{\tikz[baseline=(char.center)]{
		\node[shape=rectangle, draw, inner sep=0pt, minimum size = 5mm](char) {#1};}}
\newcommand{\csquared}[1]{\tikz[baseline=(char.center)]{ 
		\node[shape=circle, draw, inner sep=0pt, minimum size = 6mm](char) {#1};
		\node[shape=rectangle, draw, inner sep=0pt, minimum size = 5mm] at (char.center) {#1};}}

\usepackage{xcolor}

\title{Pairwise efficiency and monotonicity imply Pareto efficiency in (probabilistic) object allocation\thanks{We gratefully acknowledge financial support from the Swiss National Science Foundation (SNSF) through Project 100018-212311. We also thank Di Feng, William Thomson, Utku Ünver, the associate editor, and anonymous reviewers for their helpful comments and suggestions.}}
\author{Tom Demeulemeester\thanks{School of Economics and Business, Maastricht University, 6200 MD Maastricht, The Netherlands; \textit{e-mail}: \href{mailto:tom.demeulemeester@maastrichtuniversity.nl}{\tt tom.demeulemeester@maastrichtuniversity.nl}.} \and Bettina Klaus\thanks{\textit{Corresponding author}: Faculty of Business and Economics, University of Lausanne, 1015 Lausanne, Switzerland; \textit{e-mail}: \href{mailto:bettina.klaus@unil.ch}{\tt bettina.klaus@unil.ch}.}}
\date{\today}

\begin{document}
	\maketitle
	
	\begin{abstract}
		We consider object allocation problems with capacities where objects have to be assigned to agents. We show that a \textsl{probabilistically monotonic} lottery rule satisfies \textsl{ex-post Pareto efficiency} if and only if it satisfies \textsl{ex-post pairwise efficiency} and \textsl{ex-post non-wastefulness}. This result allows us to strengthen various existing characterization results, both for lottery rules and for deterministic rules, by replacing \textsl{(ex-post) Pareto efficiency} with \textsl{(ex-post) pairwise efficiency} and \textsl{(ex-post) non-wastefulness}, e.g., for characterizations of the Random Serial Dictatorship rule \citep{basteck2024axiomatization}, Trading Cycles rules \citep{pycia2017incentive}, and Hierarchical Exchange rules \citep{papai2000strategyproof}.\medskip
		
\medskip
		
	\end{abstract}
	
	\pagebreak

	\section{Introduction}\label{sec:intro}
	We study general object allocation problems without monetary transfers that are determined by a set of agents, a set of objects with capacities, and agents' strict preferences over objects \citep[see, e.g.,][]{abdulkadirouglu1998random,basteck2024axiomatization}.  An outcome for such a problem is a matching that, respecting capacity constraints, assigns an object to each agent.
	Our setting models, for example, the assignment of tasks to workers, or of dormitory housing to students \citep[see][for an overview of applications]{biro2017applications}.\medskip
	
	\textit{Pareto efficiency} is a key desideratum in many economic settings: an outcome is \textsl{Pareto efficient} if no agent can be made better off without making other agents worse off. In a simple object allocation model where agents have strict preferences and the number of objects equals the number of agents, this means that a matching violates \textsl{Pareto efficiency} if and only if a group of agents can trade allotments such that all members are better off. Identifying and implementing Pareto improving reallocations involving many agents, however, requires detailed knowledge of agents' preferences and a high level of coordination in decentralized environments.
	A natural question is therefore whether one really needs to rule out all Pareto improving reallocations in order to obtain efficient outcomes, or whether it is already sufficient to rule out simpler improvements, such as mutually beneficial swaps between only two agents.\medskip
	
	We therefore consider the weaker notion of \textsl{pairwise efficiency} as recently studied by \citet{ekici2024pair} and \citet{EKICI2024111459} for Shapley--Scarf housing markets \citep{shapley1974}.\footnote{Shapley--Scarf housing markets are object allocation problems where each agent owns, and wishes to consume,  exactly one object (or house).}  A matching is \textit{pairwise efficient} if and only if no pair of agents can trade, or swap, allotments such that both agents are better off. \textsl{Pairwise efficiency} can be interpreted as a local stability requirement. Bilateral swaps are the most elementary mutually beneficial deviations from a given matching and correspond to blocking pairs. In many allocation environments, Pareto improvements can be represented as trading cycles, and bilateral swaps correspond to the smallest such cycles. Bilateral improvements require only local preference information and coordination between two agents, whereas Pareto improving reallocations may involve longer improvement cycles and larger coalitions, and are therefore more difficult to identify and implement in decentralized environments. From this perspective, \textsl{pairwise efficiency} rules out the simplest and most plausible deviations from a given matching and can be viewed as a natural minimal efficiency or stability requirement.\medskip

	The focus on \textsl{pairwise efficiency} is also motivated by the classical exchange-economy literature, where bilateral trade plays a central role. In markets that allow for monetary transfers, \citet{Feldman1973bilateral} shows that, under fairly general conditions, a \textsl{pairwise efficient} allocation is also \textsl{Pareto efficient} if it assigns each trader a positive quantity of a common good (``money''), for which each trader has positive marginal utility. Alternatively, \citet{rader1968pairwise} shows that \textsl{pairwise efficiency} implies \textsl{Pareto efficiency} in barter economies in the presence of a broker,  i.e., a trader holding positive quantities of all goods. More generally, \cite{goldman1989pairwise} study necessary and sufficient conditions under which $t$-wise trading, involving $t\geq2$ agents, suffices to reach a \textsl{Pareto efficient} outcome. We refer the interested reader to \citet[][Section~3]{ostroy1990transactions} for an overview of this literature concerned with the classical exchange-economy model with monetary transfers.\medskip
	
	In markets without continuous transfers, however, little is known about the conditions under which \textsl{pairwise efficiency} and \textsl{Pareto efficiency} coincide. For Shapley--Scarf housing markets, \citet{ekici2024pair} and \citet{EKICI2024111459} characterize the Top Trading Cycles rule  as the only rule satisfying \textsl{pairwise efficiency}, \textsl{strategy-proofness}, and \textsl{individual rationality}, thus strengthening \citeauthor{ma1994}'s (\citeyear{ma1994}) classical characterization that uses \textsl{Pareto efficiency} instead of \textsl{pairwise efficiency}. In a recent paper on Shapley--Scarf housing markets, \citet{mandal2025efficient} shows that on the restricted domains of single-peaked or single-dipped preferences, \textsl{pairwise efficiency} and \textsl{Pareto efficiency} are equivalent. He also shows that these are maximal domains for this equivalence, implying that additional properties are needed for the equivalence of \textsl{pairwise efficiency} and \textsl{Pareto efficiency} to hold on larger preference domains, e.g., on the domain of strict preferences that we consider.\medskip
	
	Our main contribution is to establish equivalence results for \textsl{pairwise efficiency} and \textsl{Pareto efficiency} in general object allocation models that allow for objects with capacities, possibly with more objects than agents. In this general model, a matching also violates \textsl{Pareto efficiency} if it is \emph{wasteful}, i.e., if there exists an agent who prefers an object with remaining capacity to her allotment. We show that, on the strict preference domain, \textsl{pairwise efficiency} and \textsl{non-wastefulness} imply \textsl{Pareto efficiency} for any rule satisfying an additional \textsl{monotonicity} property. In particular, ruling out individual moves to objects with remaining capacity and mutually beneficial bilateral swaps is already sufficient to guarantee \textsl{Pareto efficiency}. Hence, a local stability requirement turns out to be strong enough to imply global efficiency.

We first show our result for the more general setting of \emph{probabilistic} object allocation problems. To satisfy even minimal fairness notions, such as the equal treatment of agents with identical preferences, some form of randomization is required. While a \emph{deterministic rule} outputs a single deterministic matching for every instance, a \emph{lottery rule} outputs a probability distribution over deterministic matchings.\medskip
	
	In this context, we show that \textsl{ex-post pairwise efficiency} and \textsl{ex-post non-wastefulness} imply  \textsl{ex-post Pareto efficiency} for any lottery rule satisfying \textsl{probabilistic monotonicity} as introduced by \citet{basteck2024axiomatization} (Theorem~\ref{th:main}). From a normative perspective, \textsl{probabilistic monotonicity} can be interpreted as requiring that if a matching becomes weakly more preferred by all agents, then the rule should not reduce the probability of selecting that matching. In this sense, the property ensures that improvements in the collective ranking of a matching are monotonically reflected in the probabilities assigned by the rule.  As a direct corollary (Corollary~\ref{cor:det_rules}), we obtain a similar equivalence result for deterministic rules by replacing \textsl{probabilistic monotonicity} with \textsl{Maskin monotonicity} (or \textsl{pairwise / group strategy-proofness}), and by replacing \textsl{ex-post non-wastefulness} with \textsl{non-wastefulness}. Additionally, for situations where there are fewer real objects than agents, there is a practically important domain restriction where all agents prefer being assigned to a real object to remaining unassigned, i.e., receiving a null object. For this restricted domain, we show that our deterministic result continues to hold (Theorem~\ref{th:restricted}).\medskip
	
	Our results also allow us to strengthen several seminal characterization results in the literature by replacing \textsl{Pareto efficiency} with \textsl{pairwise efficiency} and \textsl{non-wastefulness}, both for lottery rules and for deterministic rules. In Section~\ref{sec:previous_results}, we elaborate on how we can strengthen characterization results of the Random Serial Dictatorship rule \citep[][Theorem~1]{basteck2024axiomatization}, Trading Cycles rules \citep[][Theorems~1 and 8]{pycia2017incentive}, Hierarchical Exchange rules \citep[][main theorem]{papai2000strategyproof}, and Priority-Trees--Augmented Top Trading Cycles rules \citep[][Theorem~1]{ishida2025group}.\medskip
	
	Our paper proceeds as follows. In Section~\ref{sec:model} we introduce our object allocation model, as well as deterministic / lottery rules and their properties. In Section~\ref{sec:results}, we establish our equivalence results for \textsl{pairwise efficiency} and \textsl{Pareto efficiency} for deterministic / lottery rules (Theorem~\ref{th:main}, Corollary~\ref{cor:det_rules}, and Theorem~\ref{th:restricted}). Section~\ref{sec:previous_results} applies these results to strengthen existing characterization results (Corollaries~\ref{cor:Basteck}, \ref{cor:PyciaUnver}, \ref{cor:PyciaUnver_relaxed}, and \ref{cor:Papai}). Section~\ref{sec:conclusion} concludes and states some open questions.

	\section{The model, rules, and properties}\label{sec:model}
	
	We consider object allocation problems without transfers and with capacities where each agent wishes to consume exactly one object. 	
Let $N=\{1,\ldots,n\}$ denote a finite set of \emph{agents}, and let $O=\{o_1,\ldots,o_k\}$ denote a finite set of indivisible \emph{objects} (or object types). Each object $o\in O$ has a \emph{capacity} $q_o$, which represents the number of available copies of object $o$. We assume that each agent can receive an object, i.e., $\sum_{o\in O} q_o \geq n$. Note that the set of objects $O$ can include a null object $\varnothing$, allowing for the possibility that agents remain unassigned.\footnote{Throughout, $\varnothing$ denotes the null object, whereas $\emptyset$ denotes the empty set.}\medskip
	
	A \emph{(deterministic) matching}~$\mu$ assigns each agent $i\in N$ to exactly one object $\mu_i\in O$ while respecting the capacities of the objects, i.e., $\mu:N\to O$ is a mapping satisfying, for each $o\in O$, $|\{i\in N: \mu_i = o\}| \leq q_o$. Given a matching $\mu$, for each $i\in N$ we refer to $\mu_i$ as agent~$i$'s \textit{allotment}.  Denote the set of all matchings by $\M$.\medskip
	
	Each agent $i\in N$ has strict \emph{preferences} over all of the objects in $O$, which we represent by a linear order $\mathbin{R_i}$.\footnote{Preferences $R_i$ are a linear order if they are \emph{complete}, \emph{antisymmetric}, and \emph{transitive}. Preferences $R_i$ are \emph{complete} if, for any two objects $a, b\in O$, either $a\mathbin{R_i} b$, or $b\mathbin{R_i} a$; they are \emph{antisymmetric} if $a \mathbin{R_i} b$ and $b \mathbin{R_i} a$ imply $a=b$; and they are \emph{transitive} if for any three objects $a,b,c\in O$, $a \mathbin{R_i} b$ and $b\mathbin{R_i}c$ imply $a\mathbin{R_i}c$.} The notation $ o\mathbin{R_i} o'$ reflects that agent~$i$ \emph{weakly prefers} object $o$ to object $o'$. Agent~$i$ \emph{strictly prefers} object $o$ to object $o'$, denoted by $o \mathbin{P_i} o'$, if $o\mathbin{R_i}o'$ but not $o'\mathbin{R_i} o$. A \emph{preference profile} $R \equiv \left(\mathbin{R_i}\right)_{i\in N}$ contains the preferences of all agents in $N$. Given a preference profile $R$, for each agent $i\in N$, we use the standard notation $R_{-i} = (R_j)_{j\in N\setminus\{i\}}$ to denote the list of all agents' preferences, except for agent~$i$'s preferences. For each subset of agents $S\subseteq N$ we define $R_S=(R_i)_{i\in S}$ and $R_{-S}=(R_i)_{i\in N\setminus S}$ to be lists of preferences of the members of sets $S$ and $N\setminus S$, respectively.
	Let $\mathcal{R}$ denote the set of all strict preferences, and denote the domain of all preference profiles by $\mathcal{R}^N$. Since the sets of agents and objects (with corresponding capacities) are fixed, an instance of the \emph{object allocation problem} is represented by its preference profile $R\in \mathcal{R}^N$.\medskip
	
	Note that there are no consumption externalities and thus, at a matching $\mu$, each agent~$i$ only cares about her own allotment $\mu_i$. Hence, agent~$i$'s preferences over objects / allotments can easily be extended to preferences over matchings. With a slight abuse of notation, we use the same notation $R_i$ to denote agent~$i$'s preferences over matchings as well, that is, for each agent~$i\in N$ and any two matchings $\mu,\nu\in \M$, $\mu\mathbin{R_i}\nu$ if and only if $\mu_i\mathbin{R_i} \nu_i$.\footnote{Note that when extending strict preferences over allotments to preferences over matchings without consumption externalities, strictness is lost because an agent is indifferent between any two matchings at which she receives the same allotment.}\medskip
	
	A \emph{deterministic rule} $f:\R^N \to \M$ maps each preference profile to a (deterministic) matching. A well-known example of a deterministic rule is the \emph{Serial Dictatorship} (SD) rule, which takes a fixed order of the agents in $N$ as an input, and sequentially assigns the agents in that order to their most preferred \emph{unassigned} / \emph{remaining} object in $O$. However, any deterministic SD rule violates even very minimal fairness notions, such as the equal treatment of agents who have identical preferences. A common method to circumvent this issue is to select an order $\sigma$ of the agents in $N$ \emph{uniformly at random}, and then consider the associated expected SD rule outcome. We refer to this method as the \emph{Random Serial Dictatorship} (RSD) rule \citep[see, e.g.,][]{abdulkadirouglu1998random,bogomolnaiamoulin2001}. Note that, rather than returning a single matching, the RSD rule outputs a \emph{probability distribution} over matchings. Let $\Theta$ denote the set of all orders of the agents in $N$, with $|\Theta|=n!$. For a given preference profile $R\in\R^N$ and an order $\sigma\in \Theta$, let $\SD(\sigma, R)$ denote the matching that is obtained by running SD with order $\sigma$ on preference profile~$R$. The outcome of the RSD rule for preference profile~$R$, denoted by $\RSD(R)$, is the uniform lottery over all deterministic SD matchings, given by
	\[
	\RSD(R) = \frac{1}{n!}\sum_{\sigma\in \Theta}\SD(\sigma, R).
	\]
	
	In general, let $\Delta\M$ represent the set of all probability distributions over the matchings in $\M$. Each element in $\Delta \M$ is referred to as a \emph{lottery}.\footnote{While an element of $\Delta \M$ is sometimes also referred to as a \emph{random assignment} \citep[e.g.,][]{basteck2024axiomatization}, we use the term \emph{lottery} to distinguish between the probability distribution over matchings and the matrix containing the assignment probabilities of the agents to the objects when aggregating the lottery outcomes.} A \emph{lottery rule} $\varphi: \R^N \to \Delta \M$ maps each preference profile to a lottery. We denote by $\varphi_{\mu}(R)$ the weight of a matching $\mu\in \M$ in the lottery $\varphi(R)$. The \emph{support} of a lottery $\varphi(R)$ is the set of matchings $\mu\in\M$ with a strictly positive weight $\varphi_\mu(R)>0$. ``Deterministic preferences'' over objects can be extended to preferences over lotteries in various ways (e.g., using von Neumann--Morgenstern utility representations or stochastic dominance); since for our main results, we do not need to work with preferences over lottery outcomes, we refer the interested reader to Appendix~\ref{ap:SD-notions}.\medskip
	
	We next introduce the relevant properties of deterministic and lottery rules for our main result. We start by introducing a standard efficiency property for object allocation problems:
	
	\begin{definition}[\textbf{Pareto efficiency}]\ \\
		Given a preference profile $R\in\R^N$, a matching $\nu\in\M$ \emph{Pareto dominates} a matching $\mu\in\M$ if, for each agent $i\in N$, $\nu_i \mathbin{R_i} \mu_i$, and for at least one agent $j\in N$, $\nu_j \mathbin{P_j} \mu_j$.
		A matching $\mu\in\M$ is \emph{Pareto efficient} if there is no matching $\nu\in\M$ that Pareto dominates it.  \\A deterministic rule is \textit{Pareto efficient} if for each preference profile in $\R^N$, it assigns a \textsl{Pareto efficient} matching.
		A lottery rule is \textit{ex-post Pareto efficient} if for each preference profile in $\R^N$, it assigns a lottery with only \textsl{Pareto efficient} matchings in its support.
	\end{definition}

Note that a matching is \textsl{Pareto efficient} if and only if it is the outcome of a Serial Dictatorship rule for some order of the agents \citep{abdulkadirouglu1998random,bogomolnaia2025quotas}.\footnote{\cite{bogomolnaia2025quotas} studies a more general object allocation setting in which agents can consume multiple objects according to individual quotas and can have indifferences in their preferences. \citet[][Theorem~2]{bogomolnaia2025quotas} shows for this setting that an outcome is \textsl{Pareto efficient} if and only if it is the result of a generalized Serial Dictatorship procedure. Our setting is a special case of \citeauthor{bogomolnaia2025quotas}'s, in which all agents have a quota of one, and are indifferent between copies of the same object.}\medskip

The next property is implied by \textsl{Pareto efficiency} and models that no agent prefers an object with remaining capacity to her allotment.
	
	\begin{definition}[\textbf{Non-wastefulness}]\ \\
		Given a preference profile $R\in\R^N$, a matching $\mu\in \M$ is \emph{non-wasteful} if there does not exist an agent $i\in N$ and an object $o\in O$ such that $o \mathbin{P_i} \mu_i$ and $|\{j\in N: \mu_j = o\}| < q_o$.\\ A deterministic rule is \textit{non-wasteful} if for each preference profile in $\R^N$, it assigns a \textsl{non-wasteful} matching.	A lottery rule is \textit{ex-post non-wasteful} if for each preference profile in $\R^N$, it assigns a lottery with only \textsl{non-wasteful} matchings in its support.
	\end{definition}
	
	It is well-known that a matching is \textsl{Pareto efficient} if and only if it is \textsl{non-wasteful} and agents do not want to exchange their allotments through a so-called improvement cycle.\footnote{An \emph{improvement cycle} is an ordered sequence of agents and allotments such that each agent prefers and receives the allotment of the next agent in the sequence, and the last agent prefers and receives the allotment of the first agent.} In practice, these improvement cycles might be long and, particularly in decentralized environments, hard to detect and implement by the agents.\medskip

The following property relaxes \textsl{Pareto efficiency} by only excluding beneficial swaps between pairs of agents. Since \textsl{pairwise efficiency} rules out profitable bilateral swaps between two agents, it can be interpreted as a local stability requirement for matchings.
	
	\begin{definition}[\textbf{Pairwise efficiency}]\ \\
		Given a preference profile $R\in\R^N$, a matching $\mu\in \M$ is \emph{pairwise efficient} if there is no pair of agents $i,j \in N$ such that $\mu_j \mathbin{P_i} \mu_i$ and $\mu_i \mathbin{P_j} \mu_j$.\\
		A deterministic rule is \textit{pairwise efficient} if for each preference profile in $\R^N$, it assigns a \textsl{pairwise efficient} matching.
		A lottery rule is \textit{ex-post pairwise efficient} if for each preference profile in $\R^N$, it assigns a lottery with only \textsl{pairwise efficient} matchings in its support.
	\end{definition}
	
	\citet{ekici2024pair} introduces \textsl{pairwise efficiency} for classical Shapley--Scarf housing markets with strict preferences and strengthens a characterization of the well-known Top Trading Cycles (TTC) rule\footnote{\citet{shapley1974} introduce the top trading algorithm that the TTC rule is based on; however, they attribute it to David Gale \citep[see][]{Scarf2009MyIntroductionTTC}.} by  \citet{ma1994} by weakening \textsl{Pareto efficiency} to \textsl{pairwise efficiency}.\medskip

	Note that \textsl{pairwise efficiency} does not imply \textsl{non-wastefulness}, whereas \textsl{Pareto efficiency} does.\medskip
	
	The next three properties are incentive properties for deterministic rules that model that no agent / no pair of agents / no subset of agents can benefit from misrepresenting her~/~their preferences.
	
	\begin{definition}[\textbf{Strategy-proofness}]\ \\
		A deterministic rule $f$  is \textit{strategy-proof} if for each $R \in \mathcal{R}^N$, each agent $i\in N$, and each preference relation $R'_i\in \mathcal{R}$, $f_i(R_i,R_{-i}) \mathbin{R_i} f_i(R'_i,R_{-i})$, i.e., \textit{no agent~$i$ can manipulate rule $f$ at $R$ via $R'_i$}.
	\end{definition}
	
	\begin{definition}[\textbf{Pairwise strategy-proofness}]\ \\
		A deterministic rule $f$  is \textit{pairwise strategy-proof} if for each $R \in \mathcal{R}^N$, there is no pair of agents $i,j\in N$ (possibly, $i=j$) with preferences $R'_i,R'_j\in\mathcal{R}$ such that $f_i(R'_i,R'_j,R_{-\{i,j\}}) \mathbin{P_i} f_i(R)$ and $f_j(R'_i,R'_j,R_{-\{i,j\}}) \mathbin{R_j} f_j(R)$, i.e., \textit{no pair of agents $\{i,j\}$ can manipulate rule $f$ at $R$ via $(R'_i,R'_j)$}.
	\end{definition}
	
	\begin{definition}[\textbf{Group strategy-proofness}]\ \\	
		A deterministic rule $f$  is \textit{group strategy-proof} if for each $R \in \mathcal{R}^N$, there is no group of agents $S\subseteq N$ with preferences $R'_S=(R'_i)_{i\in S}\in \mathcal{R}^S$ such that for each $i\in S$, $f_i(R'_S,R_{-S}) \mathbin{R_i} f_i(R)$, and for some $j\in S$, $f_j(R'_S,R_{-S}) \mathbin{P_j} f_j(R)$, i.e., \textit{no group of agents $S$ can manipulate rule $f$ at $R$ via~$R'_S$}.
	\end{definition}
	
	\textsl{Group strategy-proofness} implies \textsl{pairwise strategy-proofness}, and \textsl{pairwise strategy-proofness} implies \textsl{strategy-proofness}. \medskip
	
	Next, we consider a well-known property for deterministic rules that restricts each agent's influence: no agent can influence the other agents' allotments by reporting alternative preferences that do not affect her own allotment.\medskip
	
	\begin{definition}[\textbf{Non-bossiness}]\ \\
		A deterministic rule $f$ is \textit{non-bossy} if
		for each $R \in \mathcal{R}^N$, each agent $i\in N$, and each $R'_i\in \mathcal{R}$, $f_i(R_i,R_{-i}) =f_i(R'_i,R_{-i})$ implies $f(R_i,R_{-i}) =f(R'_i,R_{-i})$.
	\end{definition}
	
	\textsl{Group strategy-proofness} implies \textsl{non-bossiness}. Furthermore, it is well-known that for object allocation problems with strict preferences and unit capacities, a deterministic rule is \textsl{group strategy-proof} if and only if it is \textsl{pairwise strategy-proof} \citep{alva2017}, and if and only if it is  \textsl{strategy-proof} and \textsl{non-bossy} \citep{papai2000strategyproof,alva2017}.\medskip

	We next introduce the well-known property of \textit{(Maskin) monotonicity} for deterministic rules, which requires that if a matching is chosen, then that matching will still be chosen if, loosely speaking, each agent shifts her allotment at that matching up in her preferences (more precisely, no agent ranks any object above her corresponding allotment that was previously ranked below it).\medskip
	
	Let $i\in N$. Given preferences $R_i\in \mathcal{R}$ and an object $o\in O$, let $L(o,R_i)=\{o'\in O : o\mathbin{R_i}o'\}$ be the \textit{lower contour set of $R_i$ at $o$}. Preference relation $R'_i$ is a \textit{monotonic transformation of $R_i$ at $o$} if $L(o,R_i)\subseteq L(o,R'_i)$.
	Similarly, given a preference profile $R\in \mathcal{R}^N$ and a matching $\mu$, a preference profile $R'\in \mathcal{R}^N$ is a \textit{monotonic transformation of $R$ at $\mu$} if for each $i\in N$, $R'_i$ is a monotonic transformation of $R_i$ at $\mu_i$.
	
	\begin{definition}[\textbf{(Deterministic Maskin) monotonicity}]\ \\
		A deterministic rule $f$ is \textit{(deterministically Maskin) monotonic} if for each $R\in \mathcal{R}^N$ and for each monotonic transformation $R'\in \mathcal{R}^N$ of $R$ at $f(R)$, $f(R')=f(R)$.
	\end{definition}
	
	It is well-known that for object allocation problems with strict preferences and unit capacities, a deterministic rule is \textsl{group strategy-proof} if and only if it is \textsl{monotonic} \citep{takamiya2001coalition,alva2017}.\medskip
	
	For later reference, we extend the above-mentioned relationships between properties from the model with unit capacities to our model with arbitrary capacities.
	
	\begin{proposition}
		The following properties are equivalent for deterministic rules:\vspace{-0.3cm}
		\begin{itemize}
			\item \textsl{group strategy-proofness},\vspace{-0.3cm}
			\item \textsl{pairwise strategy-proofness},\vspace{-0.3cm}
			\item \textsl{strategy-proofness} and \textsl{non-bossiness},\vspace{-0.3cm}
			\item \textsl{monotonicity}.\vspace{-0.3cm}
		\end{itemize}\label{prop:equivalences}
	\end{proposition}
	
	\begin{proof}[\textbf{Proof}]The extension of the model from unit capacities to arbitrary capacities does not change the proofs of \citet{alva2017}, \citet{papai2000strategyproof}, and \citet{takamiya2001coalition}. For instance, the proof of \citet[][Lemma~1]{papai2000strategyproof} establishes the equivalence between \textsl{group strategy-proofness} and [\textsl{strategy-proofness} and \textsl{non-bossiness}]; the proof of \citet[][Theorem~3.1]{takamiya2001coalition} establishes the equivalence between \textsl{group strategy-proofness} and \textsl{monotonicity}. Alternatively, the proof of \citet[][Theorem~1]{alva2017} can be used.\end{proof}

Deterministic \textsl{Maskin monotonicity} was introduced by \citet{maskin1999nash} in the context of Nash implementation. There, the condition serves as an invariance requirement: if an outcome is selected at a given preference profile and agents' preferences change in a way that preserves that outcome as at least as desirable as before (preferences are a monotonic transformation at that outcome), then the outcome should remain selected. The property subsequently became widely known as \textsl{Maskin monotonicity} and has played a central role in implementation theory \citep[see, e.g.,][]{moore1990nash}. 

As emphasized by \citet{thomsonSCW2001}, however, the substantive content of \textsl{Maskin monotonicity} can be understood as an invariance requirement---the selected outcome does not change under the admissible class of \textit{monotonic} preference transformations---rather than as a requirement that outcomes improve, or even move, in any particular direction. This invariance is consistent with a monotonicity reading: since the selected matching does not change, the (degenerate) probability assigned to it trivially weakly increases. The distinction only becomes substantive once we move to lottery rules, where probabilities can vary continuously over the unit interval rather than being restricted to the values 0 and 1.

\textit{Probabilistic monotonicity}, as proposed by  \citet{basteck2024axiomatization}, extends Maskin's property to lottery rules in a way that introduces a monotonicity requirement at the level of probabilities. When a deterministic matching $\mu$ becomes weakly more preferred by all agents---i.e., under a monotonic transformation at $\mu$---\textsl{probabilistic monotonicity} requires that the probability assigned to $\mu$ should not decrease. Thus, while deterministic \textsl{Maskin monotonicity} can be interpreted as an invariance property with respect to selected outcomes, for \textsl{probabilistic monotonicity}, the invariance and monotonicity interpretations diverge in a meaningful way.

	\begin{definition}[\textbf{Probabilistic (Maskin) monotonicity}]\ \\
		A lottery rule $\varphi$ satisfies \emph{probabilistic (Maskin) monotonicity} if, for each matching $\mu \in \M$, for each $R\in \mathcal{R}^N$, and for each monotonic transformation $R'\in \mathcal{R}^N$ of $R$ at $\mu$, $\varphi_\mu(R') \geq \varphi_{\mu}(R)$.
	\end{definition}
	
	It is easy to see that \textsl{probabilistic monotonicity} implies its deterministic version: Consider a lottery rule $\varphi$, and a preference profile $R\in\R^N$ such that $\varphi(R)=\mu$, i.e., $\varphi$ only gives a non-zero weight to a single matching~$\mu$. Then, for any preference profile $R'\in\R^N$ that is a monotonic transformation of $R$ at $\mu$, \textsl{probabilistic monotonicity} dictates that $\varphi_\mu(R') \geq \varphi_\mu(R) = 1$. Hence, $\varphi_\mu(R')=1$ and (deterministic) \textsl{monotonicity} is satisfied.\medskip
	
In Appendix~\ref{ap:exGSP}, we introduce an ex-post version of \textsl{group strategy-proofness} for the subset of lottery rules that are convex combinations of deterministic rules (so-called \textit{convex lottery rules}). We then show that for this class of lottery rules, \textsl{ex-post group strategy-proofness} is strictly stronger than \textsl{probabilistic monotonicity}.\medskip

\citet{basteck2024axiomatization} shows that probabilistic monotonicity implies \emph{sd-strategy-proofness} (see Appendix~\ref{ap:SD-notions}).\footnote{The abbreviation \emph{sd} stands for the \emph{s}tochastic \emph{d}ominance extension of preferences over objects  to lotteries over matchings \citep{bogomolnaiamoulin2001}. \citet[][Footnote~16]{basteck2024axiomatization} also explains that \textsl{sd-strategy-proofness} is equivalent to the requirement that for any von Neumann--Morgenstern utility function compatible with a given ordinal ranking of objects, submitting the true ordinal ranking maximizes an agent's expected utility.} In contrast to Proposition~\ref{prop:equivalences}, which establishes the equivalence of \textsl{monotonicity} and \textsl{group strategy-proofness} for deterministic rules, \textsl{probabilistic monotonicity} is logically independent of \emph{(weak) sd-group strategy-proofness} (see Appendix~\ref{ap:SD-notions} for the definition of \textsl{(weak) sd-group strategy-proofness} and examples showing independence).\footnote{Moreover, \citet[][Theorem~1]{basteck2024axiomatization} shows within the proof of his main result that \textsl{probabilistic monotonicity} implies \emph{weak object-wise non-bossiness}: a lottery rule satisfies \emph{weak object-wise non-bossiness} if whenever an agent swaps two adjacently ranked objects such that the total probability of receiving some object does not change, then all matchings in which the agent receives that object should be selected with the same weight in the lottery \citep[see][for a formal definition]{basteck2024axiomatization}.}\label{footnote:BasteckNB}

\section{Main results}\label{sec:results}

	While \textsl{pairwise efficiency} is a strong relaxation of \textsl{Pareto efficiency}, we show that for lottery rules satisfying \textsl{probabilistic monotonicity}, when the sum of capacities exceeds the number of agents, \textsl{ex-post pairwise efficiency} together with \textsl{ex-post non-wastefulness} is equivalent to \textsl{ex-post Pareto efficiency}. When the sum of capacities equals the number of agents, the result can be simplified by dropping the requirement of \textsl{ex-post non-wastefulness}.
	
	\begin{theorem}[\textbf{Equivalence of \textsl{ex-post pairwise} and \textsl{ex-post Pareto efficiency} on $\bm{\R^N}$}]\label{th:main}
		Let $N=\{1,\ldots,n\}$ and $O=\{o_1,\ldots,o_k\}$. Let $\varphi$ be a lottery rule on $\R^N$ that satisfies \textsl{probabilistic monotonicity}.
		\begin{itemize}
			\item[\textbf{\emph{(a)}}] If $\sum_{o\in O} q_o > n$, then $\varphi$ is \textsl{ex-post pairwise efficient} and \textsl{ex-post non-wasteful} if and only if it is \textsl{ex-post Pareto efficient}.
			\item[\textbf{\emph{(b)}}] Otherwise, $\sum_{o\in O} q_o = n$, and $\varphi$ is \textsl{ex-post pairwise efficient} if and only if it is \textsl{ex-post Pareto efficient}.
		\end{itemize}
	\end{theorem}
	
	When $\sum_{o\in O} q_o = n$, the definition of a matching implies \textsl{non-wastefulness}. The proof for Case~(a) will therefore apply to Case~(b) without explicitly requiring \textsl{non-wastefulness}.
	
	\begin{proof}[\textbf{Proof of Theorem~\ref{th:main}, Case~(a)}]\ \\ The ``if'' direction follows directly from \textsl{ex-post pairwise efficiency} and \textsl{ex-post non-wastefulness} being a relaxation of \textsl{ex-post Pareto efficiency}.\medskip
		
		To prove the ``only if'' direction, consider a lottery rule $\varphi$ that satisfies \textsl{probabilistic monotonicity}, \textsl{ex-post pairwise efficiency}, and \textsl{ex-post non-wastefulness}. Suppose, by contradiction, that $\varphi$ is not \textsl{ex-post Pareto efficient}. Hence, there exists a preference profile $R\in\R^N$ and a matching $\mu\in\M$ in the support of $\varphi(R)$ such that $\mu$ is not \textsl{Pareto efficient}. Let $\nu\in\M$ denote a matching that Pareto dominates $\mu$. As $\varphi_\mu(R) > 0$, it holds that $\varphi_\nu(R)<1$. This situation is illustrated in Figure~\ref{fig:main_R}. \medskip
		
		Let $R'$ denote the preference profile that is obtained from $R$ by moving, for each agent $i\in N$, agent~$i$'s allotment $\nu_i$ to the top of her preferences without changing the ranking of other objects, i.e.,
		\begin{itemize}
			\item[\textbf{-}] for each object $o\in O$, $\nu_i \mathbin{R'_i} o$ and
			\item[\textbf{-}] for each pair of objects $o,o'\in O\setminus\{\nu_i\}$, $o\mathbin{R'_i} o'$ if and only if $o\mathbin{R_i}o'$.
		\end{itemize}
		Note that matching $\nu$ is top-ranked by all agents for preference profile $R'$.\medskip
		
		Because $\nu$ Pareto dominates $\mu$, we have that for all agents $i \in N$, $\nu_i\mathbin{R_i}\mu_i$ and the new preference profile $R'$ is a monotonic transformation of $R$ at $\mu$. Hence, by \textsl{probabilistic monotonicity}, $\varphi_\mu(R') \geq \varphi_\mu(R) > 0$, which implies $\varphi_\nu(R')<1$. Figure \ref{fig:main_R_prime} illustrates this transformation.\medskip
		
		\begin{figure}[ht!]
			
			\centering
			\subfloat[\centering Profile $R$]{
				\begin{tikzpicture}[
					dot/.style={circle, fill=black, inner sep=1pt},
					circ/.style={circle, draw=black, inner sep=0pt, minimum size=0.6cm},
					sq/.style={rectangle, draw=black, inner sep=0pt, minimum width=0.5cm, minimum height=0.5cm}
					]
					
					\foreach \x in {0,...,6} {
						\foreach \y in {0,1,2,3} {
							\node[dot] at (\x, \y) {};
						}
					}
					
					\draw[black] (3.5, -1) -- (3.5, 4.25);
					
					\draw[black] (-0.5, 3.38) -- (6.5, 3.38);
					
					\node[circ] at (0, 2) {};
					\node[sq] at (0, 1) {};
					\node at (-0.7, 1.9) {$\nu$};
					\node at (-0.7, 0.9) {$\mu$};
					
					\node[circ] at (1, 3) {};
					\node[sq] at (1, 1) {};
					
					\node[circ] at (2, 2) {};
					\node[circ] at (3, 2) {};
					\node[sq] at (2, 0) {};
					\node[sq] at (3, 1) {};
					
					\node[sq, anchor=center] at (4, 2) {};
					\node[sq, anchor=center] at (6, 3) {};
					\node[sq, anchor=center] at (5, 1) {};
					\node[circ, anchor=center] at (4, 2) {};
					\node[circ, anchor=center] at (6, 3) {};
					\node[circ, anchor=center] at (5, 1) {};
					
					\draw[black] (-0.25,1) to[out=140, in=50] (-0.5,1);
					\draw[black] (-0.3,2) to[out=140, in=50] (-0.5,2);
					
					\foreach \x/\y in {0/3, 0/2, 2/3, 3/3, 2/2, 3/2, 4/3, 6/3, 5/2} {
						\node[dot] at (\x, \y) {};
					}
					
					\foreach \x in {0,...,6} {
						\node at (\x, -0.5) {$\vdots$};
					}
					
					\node at (0,3.75) {$R_1$};
					\node at (1,3.75) {$R_2$};
					\node at (2,3.75) {$R_3$};
					\node at (3,3.75) {$R_4$};
					\node at (4,3.75) {$R_5$};
					\node at (5,3.75) {$R_6$};
					\node at (6,3.75) {$R_7$};
					
				\end{tikzpicture} \label{fig:main_R}}%
			\qquad
			\subfloat[\centering Profile $R'$]{{\begin{tikzpicture}[
						dot/.style={circle, fill=black, inner sep=1pt},
						circ/.style={circle, draw=black, inner sep=0pt, minimum size=0.6cm},
						sq/.style={rectangle, draw=black, inner sep=0pt, minimum width=0.5cm, minimum height=0.5cm}
						]
						
						\foreach \x in {0,...,6} {
							\foreach \y in {0,1,2,3} {
								\node[dot] at (\x, \y) {};
							}
						}
						
						\draw[black] (3.5, -1) -- (3.5, 4.25);
						
						\draw[black] (-0.5, 3.38) -- (6.5, 3.38);
						
						\node[circ] at (0, 3) {};
						\node[sq] at (0, 1) {};
						\node at (-0.7, 2.9) {$\nu$};
						\node at (-0.7, 0.9) {$\mu$};
						
						\node[circ] at (1, 3) {};
						\node[sq] at (1, 1) {};
						
						\node[circ] at (2, 3) {};
						\node[circ] at (3, 3) {};
						\node[sq] at (2, 0) {};
						\node[sq] at (3, 1) {};
						
						\node[sq, anchor=center] at (4, 3) {};
						\node[sq, anchor=center] at (6, 3) {};
						\node[sq, anchor=center] at (5, 3) {};
						\node[circ, anchor=center] at (4, 3) {};
						\node[circ, anchor=center] at (6, 3) {};
						\node[circ, anchor=center] at (5, 3) {};
						
						\foreach \x/\y in {0/3, 0/2, 2/3, 3/3, 2/2, 3/2, 4/3, 6/3, 5/2} {
							\node[dot] at (\x, \y) {};
						}
						
						\foreach \x in {0,...,6} {
							\node at (\x, -0.5) {$\vdots$};
						}
						
						\draw[black] (-0.25,1) to[out=140, in=50] (-0.5,1);
						\draw[black] (-0.3,3) to[out=140, in=50] (-0.5,3);
						
						\node at (0,3.75) {$R'_1$};
						\node at (1,3.75) {$R'_2$};
						\node at (2,3.75) {$R'_3$};
						\node at (3,3.75) {$R'_4$};
						\node at (4,3.75) {$R'_5$};
						\node at (5,3.75) {$R'_6$};
						\node at (6,3.75) {$R'_7$};
						
					\end{tikzpicture} \label{fig:main_R_prime}}}%
			\caption{Illustration of preference profiles $R$ and $R'$, where the agents on the left of the vertical line in each panel represent the agents involved in the improvement cycles from matching $\mu$ to~$\nu$. Matching $\nu$ is circled and matching $\mu$ is boxed.}%
			\label{fig:example}%
		\end{figure}
		
		Let $\succ: o_1 \succ o_2 \succ \ldots \succ o_{k}$ denote a common ranking of the objects in $O$. Let $\tilR$ denote the preference profile that is obtained from $R'$ by rearranging, for each agent~$i\in N$, all objects that are ranked below object $\nu_i$ according to the common ranking~$\succ$, i.e.,
		\begin{itemize}
			\item[\textbf{-}] for each object $o\in O$, $\nu_i \mathbin{\tilR_i} o$ and
			\item[\textbf{-}] for each pair of objects $o,o'\in O\setminus\{\nu_i\}$, $o\mathbin{\tilR_i}o'$ if and only if $o \succ o'$.
		\end{itemize}
		Note that matching $\nu$ is still top-ranked by all agents for preference profile $\tilR$.\medskip
		
		Because for each agent~$i\in N$, the position of $\nu_i$ remains at the top of agent~$i$'s preferences under $R'$ and $\tilR$ (and hence unchanged), $R'$ is a monotonic transformation of $\tilR$ at $\nu$ and $\tilR$ is a monotonic transformation of $R'$ at $\nu$. Hence, by \textsl{probabilistic monotonicity} (applied twice), $\varphi_\nu(\tilR)=\varphi_\nu(R')<1$.  Figure~\ref{fig:tildeR_illustration}  gives an example of  preference profile $\tilR$ for eight agents and five objects with capacities $q = (3, 2, 1, 1, 1)$.\medskip
		
		\begin{figure}[ht!]
			
			\centering
			\renewcommand{\arraystretch}{1.2}
			\begin{tabular}{cccccccc}
				$\tilR_1$ & $\tilR_2$ & $\tilR_3$ & $\tilR_4$ & $\tilR_5$ & $\tilR_6$ & $\tilR_7$ & $\tilR_8$\\ \hline
				\circled{$a_1$} & \circled{$a_2$} & \csquared{$a_1$}& \circled{$a_3$} & \circled{$a_4$ } & \circled{$a_2$} & \circled{$a_1$} & \csquared{$a_5$} \\
				\squared{$a_2$} & \squared{$a_1$} & $a_2$ & $a_1$ & $a_1$ & \squared{$a_1$} & $a_2$ & $a_1$\\
				$a_3$ & $a_3$ & $a_3$ & $a_2$ & \squared{$a_2$} & $a_3$ & \squared{$a_3$} & $a_2$\\
				$a_4$ & $a_4$ & $a_4$ & \squared{$a_4$} & $a_3$ & $a_4$ & $a_4$ & $a_3$\\
				$a_5$ & $a_5$ & $a_5$ & $a_5$ & $a_5$ & $a_5$ & $a_5$ & $a_4$
			\end{tabular}
			\caption{An example of $\tilR$ for $n=8$, $k=5$, and $q = (3, 2, 1, 1, 1)$. Matching $\nu$ is circled and matching $\tilmu$ is boxed.\label{fig:tildeR_illustration}}
		\end{figure}	
		
		Because $\varphi_\nu(\tilR) < 1$, there exists a \textsl{pairwise efficient} and \textsl{non-wasteful} matching $\tilmu$ in the support of $\varphi(\tilR)$ that is different from $\nu$. To obtain a contradiction, we will show that any \textsl{pairwise efficient} and \textsl{non-wasteful} matching must be identical to $\nu$. \medskip
		
		Denote the set of agents who rank object $o\in O$ first at preference profile $\tilR$ by $N_{o}$, i.e., $N_{o} = \{i\in N: \text{for all } o' \in O, o \mathbin{\tilR_i} o'\}$. We first observe that, for each object, the same number of copies is assigned at $\nu$ and $\tilmu$.
		
		\begin{observation}
			\label{obs1}
			For each object $o\in O$,  $|\{i \in N: \tilmu_i = o\}|=|\{i\in N:\nu_i = o\}| = |N_o|.$
		\end{observation}
		\begin{proof}[Proof of Observation~\ref{obs1}]If Observation~\ref{obs1} were not true, then there would exist an object $o$ of which fewer copies are assigned at $\tilmu$ than at $\nu$. In that case, there must exist an agent in $N_o$ (who ranked $o$ first) who received a less preferred object at $\tilmu$. At the same time, since strictly fewer copies of $o$ are assigned at $\tilmu$ than at $\nu$, $o$ has remaining capacity at $\tilmu$, i.e., $|\{i\in N: \tilmu_i = o\}| < |N_o| \leq q_o$. Hence, $\tilmu$ violates \textsl{non-wastefulness}.
		\end{proof}
		
		To show that $\tilmu$ must be identical to $\nu$, consider the following stepwise procedure, which considers the objects in reverse order of the common ranking $\succ$.\medskip
		
		\noindent\textbf{Step~1.} We show that all agents in $N_{o_k}$ are assigned to a copy of $o_k$ at $\tilmu$, i.e., for all $i\in N_{o_k}$, $\tilmu_i = o_k$. Assume, by contradiction, that there is an agent $j\in N_{o_k}$ for whom $\tilmu_j\neq o_k$.
		
		By Observation~\ref{obs1}, at $\tilmu$, $o_k$ must be assigned to an agent $t$ in $N \setminus N_{o_k}$, i.e., $\tilmu_t = o_k$ and $t$ does not rank $o_k$ first. Note that
		\begin{enumerate}[label=(\alph*)]
			\item at $\tilmu$, agent $j$ must be assigned to an object in $O\setminus\{o_k\}$ that she prefers less than $o_k$, i.e., $\underline{o_k \mathbin{\tilP_j} \tilmu_j}$;
			\item all agents who do not belong to $N_{o_k}$ rank object $o_k$ at the bottom of their preferences. Then, since $\tilmu_j \neq o_k$, $\underline{\tilmu_j \mathbin{\tilP_t} o_k =\tilmu_t}$.
		\end{enumerate}
		Hence, agents $j$ and $t$ would like to swap their allotments at $\tilmu$, which violates \textsl{pairwise efficiency}. We now proceed to Step~2.\medskip
		
		\noindent \textbf{Step~2.} We show that all agents in $N_{o_{k-1}}$ are assigned to a copy of $o_{k-1}$ at $\tilmu$, i.e., for all $i\in N_{o_{k-1}}$, $\tilmu_i = o_{k-1}$. Assume, by contradiction, that there is an agent $j\in N_{o_{k-1}}$ for whom $\tilmu_j\neq o_{k-1}$.
		
		Recall that, by Step~1, each agent $i\in N_{o_k}$ receives her favorite object $\nu_i=o_k$ and, by Observation~\ref{obs1}, no agent in $N\setminus N_{o_k}$ is assigned a copy of object $o_k$. Hence, agent $j$ cannot be assigned to $o_k$ at $\tilmu$. Thus, $\tilmu_j \not\in\{o_{k-1}, o_k\}$. Equivalently, $\tilmu_j\in\{o_1,\ldots,o_{k-2}\}$.
		
		By Observation~\ref{obs1}, at $\tilmu$, $o_{k-1}$ must be assigned to an agent $t$ in $N \setminus \{N_{o_{k-1}}\cup N_{o_{k}}\}$, i.e., $\tilmu_t = o_{k-1}$ and $t$ does not rank $o_{k-1}$ first.  Note that
		\begin{enumerate}[label=(\alph*)]
			\item at $\tilmu$, agent $j$ must be assigned to an object in $O\setminus\{o_{k-1},o_k\}=\{o_1,\ldots,o_{k-2}\}$ that she prefers less than $o_{k-1}$ and more than $o_k$, i.e., $\underline{o_{k-1} \mathbin{\tilP_j} \tilmu_j}\mathbin{\tilP_j} o_k$;
			\item all agents who do not belong to $N_{o_{k-1}}\cup N_{o_{k}}$, i.e., they belong to $N_{o_1}\cup\ldots\cup N_{o_{k-2}}$, rank objects $o_{k-1}$ and $o_k$ at the bottom of their preferences. Then, since $\tilmu_j \in\{o_1,\ldots,o_{k-2}\}$, $\underline{\tilmu_j \mathbin{\tilP_t} o_{k-1} =\tilmu_t}\mathbin{\tilP_t} o_{k}$.			
		\end{enumerate}
		Hence, agents $j$ and $t$ would like to swap their allotments at $\tilmu$, which violates \textsl{pairwise efficiency}. If $k = 2$, terminate the procedure. Otherwise, we proceed to Step~3, and so on.\medskip
		
		\noindent\textbf{Step~$\bm{\st}$.} We show that all agents in $N_{o_{k+1-\st}}$ are assigned to a copy of $o_{k+1-\st}$ at $\tilmu$, i.e., for all $i\in N_{o_{k+1-\st}}$, $\tilmu_i = o_{k+1-\st}$. Assume, by contradiction, that there is an agent $j\in N_{o_{k+1-\st}}$ for whom $\tilmu_j\neq o_{k+1-\st}$.
		
		Recall that, by Steps~$1,\ldots,\st-1$, each agent $i\in N_{o_{k+2-\st}}\cup\ldots\cup  N_{o_k}$ receives her favorite object $\nu_i\in\{o_{k+2-\st},\ldots,o_k\}$ and, by Observation~\ref{obs1}, no agent in $N_{o_1}\cup\ldots\cup N_{o_{k+1-\st}}$ is assigned a copy of an object in $\{o_{k+2-\st},\ldots,o_k\}$. Hence, agent $j$ cannot be assigned to an object in $\{o_{k+2-\st},\ldots,o_k\}$ at $\tilmu$. Thus, $\tilmu_j \not\in \{o_{k+1-\st},o_{k+2-\st},\ldots,o_k\}$. Equivalently, $\tilmu_j\in\{o_1,\ldots,o_{k-\st}\}$.
		
		By Observation~\ref{obs1}, at $\tilmu$, $o_{k+1-\st}$ must be assigned to an agent $t$ in $N_{o_1}\cup\ldots\cup N_{o_{k-\st}}$, i.e., $\tilmu_t = o_{k+1-\st}$ and $t$ does not rank $o_{k+1-\st}$ first.  Note that
		
		\begin{enumerate}[label=(\alph*)]
			\item at $\tilmu$, agent $j$ must be assigned to an object in $\{o_1,\ldots,o_{k-\st}\}$ that she prefers less than $o_{k+1-\st}$ and more than objects in $\{o_{k+2-\st},\ldots,o_k\}$, i.e., $\underline{o_{k+1-\st} \mathbin{\tilP_j} \tilmu_j}\mathbin{\tilP_j}o_{k+2-\st} \mathbin{\tilP_j}\ldots\mathbin{\tilP_j}o_k$;
			\item all agents in $N_{o_1}\cup\ldots\cup N_{o_{k-\st}}$ rank objects in $\{o_{k+2-\st},\ldots,o_k\}$ at the bottom of their preferences. Then, since $\tilmu_j\in\{o_1,\ldots,o_{k-\st}\}$, $\underline{\tilmu_j \mathbin{\tilP_t} o_{k+1-\st} =\tilmu_t}\mathbin{\tilP_t}o_{k+2-\st}\mathbin{\tilP_t} \ldots\mathbin{\tilP_t}o_{k}$.			
		\end{enumerate}
		Hence, agents $j$ and $t$ would like to swap their allotments at $\tilmu$, which violates \textsl{pairwise efficiency}. If $k=\st$, terminate the procedure. Otherwise, proceed to Step~$\st+1$.\medskip
		
		As a result of this stepwise procedure, it must hold that $\tilmu = \nu$, which yields the desired contradiction.\end{proof}
	
	As a direct corollary of Theorem~\ref{th:main}, we obtain similar equivalence conditions between \textsl{pairwise efficiency} and \textsl{Pareto efficiency} for deterministic rules.
	
	\begin{corollary}[\textbf{Equivalence of \textsl{pairwise} and \textsl{Pareto efficiency} on $\bm{\R^N}$}]\label{cor:det_rules}\ \\
		Let $N=\{1,\ldots,n\}$ and $O=\{o_1,\ldots,o_k\}$. Let $f$ be a deterministic rule on $\R^N$ that satisfies \textsl{pairwise strategy-proofness}.
		\begin{itemize}
			\item[\textbf{\emph{(a)}}] If $\sum_{o\in O} q_o > n$, then $f$ is \textsl{pairwise efficient} and \textsl{non-wasteful} if and only if it is \textsl{Pareto efficient}.
			\item[\textbf{\emph{(b)}}] Otherwise, $\sum_{o\in O} q_o = n$, and $f$ is \textsl{pairwise efficient} if and only if it is \textsl{Pareto efficient}.
		\end{itemize}
	\end{corollary}
	
	Note that in the statement of Corollary~\ref{cor:det_rules}, we use Proposition~\ref{prop:equivalences} to replace \textsl{monotonicity} with \textsl{pairwise strategy-proofness}. By doing so, we obtain a further strengthening of our main result because preventing coalitions of at most two agents
	\begin{itemize}
		\item from modifying the outcome of a deterministic rule through a misreport (\textsl{pairwise strategy-proofness}), and
		\item from improving the matching through either an individual move to an object with remaining capacity or a bilateral swap (\textsl{non-wastefulness} and \textsl{pairwise efficiency}),
	\end{itemize}
	is enough to prevent arbitrary groups of agents
	\begin{itemize}
		\item from modifying the outcome through a misreport (\textsl{group strategy-proofness}), and
		\item from improving the matching through a Pareto improving reallocation (\textsl{Pareto efficiency}).
	\end{itemize}
		
	We next show that Corollary~\ref{cor:det_rules} also holds for a commonly studied domain restriction in object allocation problems \citep[e.g.,][]{papai2000strategyproof, basteck2024axiomatization}. In object allocation problems where all real objects are desirable and agents can receive a null object $\varnothing\in O$ that represents ``not receiving a real object,'' it is natural to restrict the preference domain by assuming that the null object appears at the bottom of the agents' preferences. Let $\underline{\R}\varsubsetneq \R$ denote the set of all strict preferences such that each real object is preferred to the null object, i.e., for each $R\in \underline{\R}$, and each $o\in O\setminus\{\varnothing\}$, $o \mathbin{P} \varnothing$. Let $\underline{\R}^N\varsubsetneq\R^N$ denote the set of all preference profiles in this restricted domain. Note that Proposition~\ref{prop:equivalences} continues to hold when we replace domain $\R$ by the restricted domain $\underline{\R}$.\medskip
	
	Unfortunately, the proof of Theorem~\ref{th:main} does not directly apply to this modified setting for object allocation problems with scarcity where some agents must receive the null object, as it could involve preferences at which the null object does not appear at the bottom of an agent's preferences after monotonically transforming the preferences, thereby violating the domain restriction $\underline{\R}^N$. With Theorem~\ref{th:restricted}, we show a corresponding result for deterministic rules on $\underline{\R}^N$; for that result, we develop a different proof strategy in Appendix~\ref{app:proof_restricted} that, alas, does not extend to lottery rules.\footnote{Whether Theorem~\ref{th:main} holds for lottery rules on $\underline{\R}^N$ when there is scarcity remains an open problem.\label{footnote:openproblem}}

	\begin{theorem}[\textbf{Equivalence of \textsl{pairwise} and \textsl{Pareto efficiency} on $\bm{\underline{\R}^N}$}] \label{th:restricted}\ \\
		Let $N=\{1,\ldots,n\}$ and $O=\{\varnothing,o_1,\ldots,o_k\}$.
		Let $f$ be a deterministic rule on $\underline{\R}^N$ that satisfies \textsl{pairwise strategy-proofness}.
		\begin{itemize}
			\item[\textbf{\emph{(a)}}] If $\sum_{o\in O} q_o > n$, then $f$ is \textsl{pairwise efficient} and \textsl{non-wasteful} if and only if it is \textsl{Pareto efficient}.
			\item[\textbf{\emph{(b)}}] Otherwise, $\sum_{o\in O} q_o = n$, and $f$ is \textsl{pairwise efficient} if and only if it is \textsl{Pareto efficient}.
		\end{itemize}
	\end{theorem}
	
	Note that $\sum_{o\in O} q_o = n$ implies $q_{\varnothing}=n-\sum_{o\in O\setminus\{\varnothing\}}q_o$ and the definition of a matching implies \textsl{non-wastefulness}.\footnote{Note that $q_\varnothing = n$ is a common assumption in the object allocation literature; it allows for the ``null matching'' in which no agent receives a real object. A capacity $q_\varnothing < n$ on domain restriction $\underline{\R}^N$ can be interpreted as $\varnothing$ being an undesirable real object with limited capacity that is ranked last by all agents.} The proof for Case~(a) in Appendix~\ref{app:proof_restricted} will therefore also apply to Case~(b) without explicitly requiring \textsl{non-wastefulness}.\medskip

	\section{Strengthening previous characterization results}
	\label{sec:previous_results}
	
	The results in Section~\ref{sec:results} directly allow us to strengthen several existing results in the literature, both for lottery rules and for deterministic rules.\medskip
	
	\subsection{Characterizing Random Serial Dictatorship rules}\label{subsec:RSD}
	
	Theorem~\ref{th:main} allows us to strengthen a recent characterization of the Random Serial Dictatorship rule by \citet[][Theorem~1]{basteck2024axiomatization}. He uses the following additional property in his result. A lottery rule satisfies \textsl{equal treatment of equals} if, for any two agents with identical preferences, the weight of a matching in the lottery is the same as the weight of the alternative matching in which the roles of these two agents are permuted.
	
	\begin{definition}[\textbf{Equal treatment of equals}]\ \\A lottery rule $\varphi$ satisfies \textsl{equal treatment of equals} if, for each preference profile $R\in\R^N$ with a pair of agents $i,j\in N$ who have identical preferences $R_i = R_j$, the following holds. Let matchings $\mu, \mu'\in \M$ be such that $\mu_i = \mu'_j$, $\mu'_i = \mu_j$, and for each $l\in N\setminus\{i,j\}$, $\mu_l = \mu'_l$. Then, $\varphi_\mu(R) = \varphi_{\mu'}(R)$.
	\end{definition}
	
	\citet[][Theorem~1]{basteck2024axiomatization} shows that the Random Serial Dictatorship rule, for the object allocation model as defined in Section~\ref{sec:model}, is characterized by \textsl{ex-post Pareto efficiency}, \textsl{probabilistic monotonicity}, and \textsl{equal treatment of equals}. Theorem~\ref{th:main} directly implies that \textsl{ex-post Pareto efficiency} can be weakened to \textsl{ex-post pairwise efficiency} and \textsl{ex-post non-wastefulness}.
	
	\begin{corollary}[\textbf{A characterization of the Random Serial Dictatorship rule on $\bm{\R^N}$}]\label{cor:Basteck}\ \\
		Let $N=\{1,\ldots,n\}$ and $O=\{o_1,\ldots,o_k\}$.  Let $\varphi$ be a lottery rule on $\R^N$.
		\begin{itemize}
			\item[\textbf{\emph{(a)}}] If $\sum_{o\in O} q_o > n$, then $\varphi$ satisfies \textsl{probabilistic monotonicity}, \textsl{ex-post pairwise efficiency}, \textsl{equal treatment of equals}, and \textsl{ex-post non-wastefulness} if and only if $\varphi$ is the Random Serial Dictatorship rule.
			\item[\textbf{\emph{(b)}}] Otherwise, $\sum_{o\in O} q_o = n$, and $\varphi$ satisfies \textsl{probabilistic monotonicity}, \textsl{ex-post pairwise efficiency}, and \textsl{equal treatment of equals} if and only if $\varphi$ is the Random Serial Dictatorship rule.
		\end{itemize}
	\end{corollary}
	
	Note that \citet[][Theorem~1]{basteck2024axiomatization} contains further characterizations of  the Random Serial Dictatorship rule that involve properties based on marginal probability comparisons (\textsl{stochastic dominance strategy-proofness}, \textsl{weak object-wise non-bossiness}, \textsl{pairwise responsiveness}), but, according to \citeauthor{basteck2024axiomatization}, ``The characterization based on probabilistic monotonicity might be seen as the most compelling on normative grounds.'' Furthermore, \citet[][Theorem~1]{basteck2024axiomatization} establishes his results for two preference profile domains, $\R^N$ and $\underline{\R}^N$. As already mentioned in Footnote~\ref{footnote:openproblem}, it is an open question whether our Theorem~\ref{th:main} holds for lottery rules on $\underline{\R}^N$; hence it is also an open question whether in  \citet[][Theorem~1]{basteck2024axiomatization}, for the restricted domain $\underline{\R}^N$, \textsl{ex-post Pareto efficiency} can be weakened to \textsl{ex-post pairwise efficiency} and \textsl{ex-post non-wastefulness}.\medskip
	
	\subsection{Strengthening characterization results for deterministic rules}\label{subsec:determ}
	Corollary~\ref{cor:det_rules} and Theorem~\ref{th:restricted} also allow us to strengthen several seminal results for deterministic object allocation problems, which were established for less general object allocation models than the one introduced in Section~\ref{sec:model}.\medskip
	
	The first important result in the literature that can be strengthened by Corollary~\ref{cor:det_rules} is a characterization result by \citet{pycia2017incentive} \citep[see also][]{bade2020random}.\footnote{\citet{bade2020random} refers to a deterministic rule that satisfies \textsl{Pareto efficiency}, \textsl{strategy-proofness}, and \textsl{non-bossiness} as \textsl{good}. Corollary~\ref{cor:det_rules} implies that what \citeauthor{bade2020random} calls \textsl{goodness} of a rule is equivalent to \textsl{pairwise efficiency}, \textsl{non-wastefulness}, \textsl{strategy-proofness}, and \textsl{non-bossiness}.} They study a model without a null object and with a unit-capacity object set that contains at least as many objects as agents. \citeauthor{pycia2017incentive} characterize the set of \textsl{Pareto efficient} and \textsl{group strategy-proof} deterministic rules as the class of so-called \emph{Trading Cycles rules}.\footnote{\citeauthor{pycia2017incentive}'s Trading Cycles rules build on the classic Top Trading Cycles (TTC) algorithm but, instead of simply owning objects, agents might have different kinds of control rights over objects (which may sequentially depend on the executed steps of the algorithm): they can own objects, or act as brokers who control how the item is passed along. At each step of the Trading Cycles algorithm, agents look at the available objects and point to the one they most prefer according to their control rights (owners can point at their best available object while brokers essentially can only point at best available objects owned by others). At the same time, each item points to the agent who currently controls it, whether by ownership or brokerage. These directed links create trading cycles that can be cleared accordingly. The agents and objects involved are then removed, and the process repeats with the remaining agents and objects.} Proposition~\ref{prop:equivalences} and Corollary~\ref{cor:det_rules} then allow us to weaken \textsl{group strategy-proofness} to \textsl{pairwise strategy-proofness} and to weaken \textsl{Pareto efficiency} to \textsl{pairwise efficiency} and \textsl{non-wastefulness} in \citet[][Theorem~1]{pycia2017incentive}, thus obtaining the following strengthened characterization of Trading Cycles rules on $\R^N$.
	
\begin{corollary}[\textbf{Characterizing Trading Cycles rules on $\bm{\R^N}$}]\ \\
		Let $N=\{1,\ldots,n\}$, $O=\{o_1,\ldots,o_k\}$, for each $o\in O$, $q_o=1$, and  $k\geq n$. Let $f$ be a deterministic rule on $\R^N$.
		\begin{itemize}
			\item[\textbf{\emph{(a)}}] If $\sum_{o\in O} q_o > n$, then $f$ satisfies \textsl{pairwise efficiency}, \textsl{pairwise strategy-proofness}, and \textsl{non-wastefulness} if and only if $f$ is a Trading Cycles rule.
			\item[\textbf{\emph{(b)}}] Otherwise, $\sum_{o\in O} q_o = n$, and $f$ satisfies \textsl{pairwise efficiency} and \textsl{pairwise strategy-proofness} if and only if $f$ is a Trading Cycles rule.
	\end{itemize}\label{cor:PyciaUnver}
	\end{corollary}
	
	In their supplementary material, \citet[][Theorem~8]{pycia2017incentive} show that their result still holds when the number of objects can be smaller than the number of agents, thus allowing for a null object.\footnote{Recall that our general model allows for a null object to be included in set $O$.} In particular, they show that their result holds on the restricted domain~$\underline{\R}^N$. Proposition~\ref{prop:equivalences} and Theorem~\ref{th:restricted} then allow us to weaken \textsl{group strategy-proofness} to \textsl{pairwise strategy-proofness} and to weaken \textsl{Pareto efficiency} to \textsl{pairwise efficiency} and \textsl{non-wastefulness} in \citet[][Theorem~8]{pycia2017incentive}, thus obtaining the following strengthened characterization of Trading Cycles rules on $\underline{\R}^N$.
	
	\begin{corollary}[\textbf{Characterizing Trading Cycles rules on $\bm{\underline{\R}^N}$}]\ \\
		Let $N=\{1,\ldots,n\}$ and $O=\{\varnothing, o_1,\ldots,o_k\}$ such that $q_\varnothing \geq n-k$ and for each $o\in O\setminus\{\varnothing\}$, $q_o=1$. Let $f$ be a deterministic rule on $\underline{\R}^N$.
		\begin{itemize}
			\item[\textbf{\emph{(a)}}] If $\sum_{o\in O} q_o > n$, then $f$ satisfies \textsl{pairwise efficiency}, \textsl{pairwise strategy-proofness}, and \textsl{non-wastefulness} if and only if $f$ is a Trading Cycles rule.
			\item[\textbf{\emph{(b)}}] Otherwise, $\sum_{o\in O} q_o = n$, and $f$ satisfies \textsl{pairwise efficiency} and \textsl{pairwise strategy-proofness} if and only if $f$ is a Trading Cycles rule.
		\end{itemize}\label{cor:PyciaUnver_relaxed}
	\end{corollary}
	
	Another classical result for object allocation problems that can be strengthened by Corollary~\ref{cor:det_rules} is a characterization by \citet[][main theorem]{papai2000strategyproof}. She studies the same model as the one described in Corollary~\ref{cor:PyciaUnver_relaxed}, with a unit-capacity object set that includes a null object (with capacity $n$) and without restriction on the  number of objects $k$ compared to the number of agents $n$. On the restricted domain $\underline{\R}^N$, \citet[][main theorem]{papai2000strategyproof} characterizes the class of Hierarchical Exchange rules, a subclass of  \citeauthor{pycia2017incentive}'s Trading Cycles rules, with the additional property of \textsl{self-enforcing--reallocation-proofness}.\footnote{\citet{papai2000strategyproof} introduces a property that excludes joint preference manipulation by two agents who plan to swap objects ex-post under the condition that the collusion changed both their allotments and is self-enforcing in the sense that neither agent changes her allotment in case she misreports while the other agent reports the truth. Formally, a deterministic rule $f$ is \textit{self-enforcing--reallocation-proof} if for each $R\in \underline{\R}^N$, there is no pair of agents $i,j\in N$ with preferences $R'_i,R'_j\in \underline{\R}$ such that $f_j(R'_i,R'_j,R_{-\{i,j\}})\mathbin{R_i}f_i(R)$ and $f_i(R'_i,R'_j,R_{-\{i,j\}})\mathbin{R_j}f_j(R),$ and  for $l=i,j$, $f_{l}(R) = f_{l}(R'_l,R_{-l})\neq f_l(R'_i,R'_j,R_{-\{i,j\}})$.} \citeauthor{papai2000strategyproof}'s Hierarchical Exchange rules can be obtained from the class of Trading Cycles rules by excluding brokerage. Proposition~\ref{prop:equivalences} and Theorem~\ref{th:restricted} then allow us to weaken \textsl{group strategy-proofness} to \textsl{pairwise strategy-proofness} and to weaken \textsl{Pareto efficiency} to \textsl{pairwise efficiency} and \textsl{non-wastefulness} in \citet[][main theorem]{papai2000strategyproof}, thus  obtaining the following strengthened characterization of Hierarchical Exchange rules on $\underline{\R}^N$.	
	
	\begin{corollary}[\textbf{Characterizing Hierarchical Exchange rules on $\bm{\underline{\R}^N}$}]\ \\
		Let $N=\{1,\ldots,n\}$, $O=\{\varnothing,o_1,\ldots,o_k\}$, $q_{\varnothing}\geq n-k$, and for each $o\in O\setminus\{\varnothing\}$, $q_o=1$. Let $f$ be a deterministic rule on $\underline{\R}^N$.
		\begin{itemize}
			\item[\textbf{\emph{(a)}}] If $\sum_{o\in O} q_o > n$, then $f$ satisfies \textsl{pairwise efficiency}, \textsl{pairwise strategy-proofness},  \textsl{non-wastefulness}, and \textsl{self-enforcing--reallocation-proofness} if and only if $f$ is a Hierarchical Exchange rule.
			\item[\textbf{\emph{(b)}}] Otherwise, $\sum_{o\in O} q_o = n$, and $f$ satisfies \textsl{pairwise efficiency}, \textsl{pairwise strategy-proofness}, and \textsl{self-enforcing--reallocation-proofness} if and only if $f$ is a Hierarchical Exchange rule.\footnote{\citet{papai2000strategyproof} states her characterization for $q_{\varnothing}=n$. Since the null object is ranked last by all agents, under \textsl{non-wastefulness} any null capacity beyond $n-k$ is never allocated; hence the characterization is identical for every $q_{\varnothing}\geq n-k$, of which \citeauthor{papai2000strategyproof}'s model ($q_{\varnothing}=n$) is a special case.}
		\end{itemize}\label{cor:Papai}
	\end{corollary}
	
	Finally, Proposition~\ref{prop:equivalences} and Corollary~\ref{cor:det_rules} / Theorem~\ref{th:restricted} can be used to strengthen characterizations of subfamilies of Trading Cycles or Hierarchical Exchange rules. For instance, \citet{ishida2025group} study object allocation problems with a specific ownership structure in which the set of agents is partitioned into groups that each own a number of objects equal to their size. \citet[][Theorem~1]{ishida2025group} then characterize the set of \emph{Priority-Trees--Augmented} Top Trading Cycles rules as the only rules satisfying \textsl{Pareto efficiency}, \textsl{group strategy-proofness}, \textsl{within-group endowments lower bounds}, and \textsl{group-wise neutrality} \citep[for the motivations and definitions of the last two properties, we refer the interested reader to][]{ishida2025group}. In the same manner as before, Proposition~\ref{prop:equivalences} and Corollary~\ref{cor:det_rules} allow us to weaken \textsl{group strategy-proofness} to \textsl{pairwise strategy-proofness} and to weaken \textsl{Pareto efficiency} to \textsl{pairwise efficiency} and \textsl{non-wastefulness} in \citet[][Theorem~1]{ishida2025group}, thus  obtaining a strengthened characterization of Priority-Trees--Augmented Top Trading Cycles rules.
	
\section{Conclusion and open questions}\label{sec:conclusion}

The central insight of our paper is that, in object allocation problems with capacities and under suitable incentive or preference responsiveness conditions, it is not necessary to impose full \textsl{Pareto efficiency} directly. It is sufficient to rule out individual moves to objects with remaining capacity and mutually beneficial bilateral swaps. In this sense, a local stability requirement turns out to be strong enough to guarantee global efficiency.

More precisely, our main result shows that for lottery rules satisfying \textsl{probabilistic monotonicity}, \textsl{ex-post pairwise efficiency} together with \textsl{ex-post non-wastefulness} is equivalent to \textsl{ex-post Pareto efficiency}. When the total capacity equals the number of agents, \textsl{ex-post non-wastefulness} is automatically satisfied, and \textsl{ex-post pairwise efficiency} alone is equivalent to \textsl{ex-post Pareto efficiency}. As a direct consequence, we obtain analogous equivalence results for deterministic rules by replacing \textsl{probabilistic monotonicity} with \textsl{group} or \textsl{pairwise strategy-proofness} (or with \textsl{strategy-proofness} and \textsl{non-bossiness}). We also show that on the restricted domain where all agents rank the null object last, the deterministic equivalence result continues to hold.

These equivalence results allow us to strengthen several existing characterization results in the literature. In particular, \textsl{Pareto efficiency} can be replaced by \textsl{pairwise efficiency} (together with \textsl{non-wastefulness} under excess capacity) in characterizations of the Random Serial Dictatorship rule, Trading Cycles rules, Hierarchical Exchange rules, and Priority-Trees--Augmented Top Trading Cycles rules.\medskip

Note that all our equivalence results rely on a preference responsiveness condition, \textsl{probabilistic monotonicity} for lottery rules, and on a group incentive condition, \textsl{group strategy-proofness} for deterministic rules. To illustrate the necessity of these conditions for our equivalence results to hold, consider the following deterministic rule, which satisfies \textsl{pairwise efficiency}, but violates \textsl{strategy-proofness} and \textsl{Pareto efficiency}.\medskip

Let \(N=\{1,2,3\}\), \(O=\{a,b,c\}\), and \(q=(1,1,1)\). Let \(\widehat R=(\widehat R_1,\widehat R_2,\widehat R_3)\in\mathcal R^N\) be the preference profile given by
\[
\widehat R_1:\ a \;\widehat P_1\; b \;\widehat P_1\; c,
\qquad
\widehat R_2:\ b \;\widehat P_2\; c \;\widehat P_2\; a,
\qquad
\widehat R_3:\ c \;\widehat P_3\; a \;\widehat P_3\; b.
\]
Let $\sigma=(1,2,3)$ and define the deterministic rule \(f\) by
\[
f(R)=
\begin{cases}
(b,c,a), & \text{if } R=\widehat R,\\
\SD(\sigma,R), & \text{if } R\neq \widehat R.
\end{cases}
\]

Rule \(f\) is \textsl{pairwise efficient}, but violates \textsl{Pareto efficiency} at profile $\widehat{R}$. Moreover, rule \(f\) is not \textsl{strategy-proof}. At profile \(\widehat R\), agent \(2\) receives object \(c\).
If instead agent \(2\) reports
\[
R_2':\ a \;P_2'\; b \;P_2'\; c,
\]
while agents \(1\) and \(3\) keep reporting \(\widehat R_1\) and \(\widehat R_3\), respectively, then \((\widehat R_1,R_2',\widehat R_3)\neq \widehat R\), and therefore
\[
f(\widehat R_1,R_2',\widehat R_3)
=
\SD((1,2,3),(\widehat R_1,R_2',\widehat R_3))
=
(a,b,c).
\]
Hence, agent \(2\) receives object \(b\). Since under agent \(2\)'s true preferences \(\widehat R_2\), \(b \;\widehat P_2\; c\), agent \(2\) can profitably manipulate \(f\). Therefore, rule \(f\) is not \textsl{strategy-proof}.\footnote{One can also formulate a lottery rule that satisfies \textsl{equal treatment of equals} and violates \textsl{probabilistic monotonicity} by assigning the same matching \((b,c,a)\) at profile \(\widehat R\) and applying the Random Serial Dictatorship rule otherwise.}\medskip

This raises the \textbf{open question} whether our results can be obtained under weaker incentive or preference responsiveness conditions. For lottery rules, it is not yet clear what the right weakening of \textsl{probabilistic monotonicity} should be; Appendices~\ref{ap:exGSP} and \ref{ap:SD-notions} discuss several ex-post and ex-ante notions, but the corresponding group incentive properties are either stronger than or logically independent of \textsl{probabilistic monotonicity}. A natural idea then would be to switch from group to individual non-manipulability. However, even for deterministic rules, we do not currently know whether our results extend to this case. In particular, we do not know whether \textsl{pairwise efficiency} and \textsl{non-wastefulness} already imply \textsl{Pareto efficiency} for every \textsl{strategy-proof} deterministic rule, nor do we have a counterexample: a deterministic rule that is \textsl{strategy-proof}, \textsl{pairwise efficient}, and  \textsl{non-wasteful} but fails to be \textsl{Pareto efficient} (and, in view of Corollary~\ref{cor:det_rules}, violates \textsl{non-bossiness}!).\medskip

Finally, another \textbf{open question} concerns the restricted domain in which all agents rank the null object last. While we obtain equivalence results for deterministic rules on this domain (Theorem~\ref{th:restricted}), we do not know whether an analogous result can be obtained for lottery rules. In particular, we do not know whether, for lottery rules on this restricted domain, \textsl{ex-post pairwise efficiency}, \textsl{ex-post non-wastefulness}, and \textsl{probabilistic monotonicity} imply \textsl{ex-post Pareto efficiency}.

	\appendix

\makeatletter
\renewcommand{\@seccntformat}[1]{Appendix~\csname the#1\endcsname:\quad}
\makeatother

\section{Ex-post group strategy-proofness}\label{ap:exGSP}

An interesting subset of lottery rules is given by those that are defined as convex combinations of deterministic rules \citep[see, e.g.,][]{bade2016fairness,bade2020random}. One example is the Random Serial Dictatorship, which is the uniformly weighted average over the set of all deterministic Serial Dictatorship rules, as defined in Section~\ref{sec:model}. By modifying the weights that are assigned to various deterministic Serial Dictatorship rules, other ``SD lottery rules'' can be constructed.  More generally, let $\mathcal{F}$ denote the set of all deterministic rules, and let $\Delta \mathcal{F}$ represent the set of all probability distributions over the deterministic rules in $\mathcal{F}$. \medskip

A \emph{convex lottery rule} $\varphi\in \Delta\mathcal{F}$ maps each preference profile to the corresponding lottery over the matchings that are assigned by the deterministic rules. To be more precise, we denote by $\varphi_f$ the weight of a deterministic rule $f\in \mathcal{F}$ in lottery $\varphi\in \Delta\mathcal{F}$. The \emph{support} of a convex lottery rule $\varphi$ is the set of deterministic rules $f \in \mathcal{F}$ with a strictly positive weight $\varphi_f>0$. Then, convex lottery rule $\varphi\in \Delta\mathcal{F}$ induces lottery rule $\varphi: \R^N \to \Delta\mathcal{M}$ such that for each $R\in\R^N$ and each $\mu\in \mathcal{M}$, $\varphi_\mu(R)=\sum_{f: f(R)= \mu} \varphi_f$. Note that the set of convex lottery rules is a strict subset of the set of all lottery rules.\medskip
	
	If a convex lottery rule only contains \textsl{group strategy-proof} deterministic rules in its support, we say that this convex lottery rule is \emph{ex-post group strategy-proof}.
	
	\begin{definition}[\textbf{Ex-post group strategy-proofness}]\ \\
		A convex lottery rule $\varphi$ is \textsl{ex-post group strategy-proof} if, for every deterministic rule $f\in\mathcal{F}$, $\varphi_f>0$ implies that $f$ is \textsl{group strategy-proof}.
	\end{definition}
	
	Similarly, a convex lottery rule is \emph{ex-post strategy-proof} if every deterministic rule in its support is \textsl{strategy-proof}.
	
	Because of the equivalence of \textsl{group strategy-proofness} and \textsl{monotonicity} for deterministic rules (Proposition~\ref{prop:equivalences}), we obtain that any \textsl{ex-post group strategy-proof} convex lottery rule satisfies \textsl{probabilistic monotonicity}.
	
	\begin{lemma}
		\label{lemma:GSP}
		If a convex lottery rule is \textsl{ex-post group strategy-proof}, then it satisfies \textsl{probabilistic monotonicity}.
	\end{lemma}

	\begin{proof}[\textbf{Proof}]
		Consider a convex lottery rule $\varphi\in \Delta\mathcal{F}$ that satisfies \textsl{ex-post group strategy-proofness}. In order to prove \textsl{probabilistic monotonicity}, let $\mu \in \M$ and  $R\in \mathcal{R}^N$, and consider a monotonic transformation $R'\in \mathcal{R}^N$ of $R$ at $\mu$. We then need to show that $\varphi_\mu(R') \geq \varphi_{\mu}(R)$.
By Proposition~\ref{prop:equivalences}, every deterministic rule $f$ in the support of $\varphi$ satisfies \textsl{monotonicity}, and $f(R)=\mu$ implies $f(R')=f(R)=\mu$. Thus,  $$\{f\in \mathcal{F}: \varphi_f > 0 \text{ and }f(R)= \mu\}\subseteq\{f\in \mathcal{F}: \varphi_f > 0 \text{ and } f(R') = \mu\}$$ and  $$\varphi_\mu(R)=\sum_{f\in \mathcal{F}: f(R)= \mu} \varphi_f\leq \sum_{f\in \mathcal{F}: f(R') = \mu} \varphi_f=\varphi_\mu(R').$$
Hence, the convex lottery rule $\varphi$ satisfies \textsl{probabilistic monotonicity}.
	\end{proof}

On the other hand, the following is an example of a lottery rule that satisfies \textsl{ex-post Pareto efficiency} and \textsl{probabilistic monotonicity}, but violates \textsl{ex-post strategy-proofness}.\footnote{A simpler example is given in \citet[][proof of Theorem~4]{PYCIA201521}: a single-agent lottery rule that assigns the agent either her most preferred or her second-most preferred object with equal probability. That lottery rule satisfies \textsl{probabilistic monotonicity} but is neither \textsl{ex-post Pareto efficient} nor \textsl{ex-post strategy-proof}.}
Let \(N=\{1,2\}\) and \(O=\{a,b,c\}\), with capacities $q_a=q_b=1$ and $q_c=2$.
Let $\sigma^1=(1,2)$ and $\sigma^2=(2,1)$ be the two Serial Dictatorship orders. Moreover, let
$\widehat R=(\widehat R_1,\widehat R_2)\in\mathcal R^N$ be the preference profile such that,
for each $i\in N$, $\widehat R_i:\ a \mathbin{\widehat P_i} b \mathbin{\widehat P_i} c$.

Define the following \textsl{Pareto efficient} deterministic rules $f^{\sigma^1},f^{\sigma^2}\in \mathcal{F}$ such that for each
$R\in\mathcal R^N$,
\[
f^{\sigma^1}(R)=
\begin{cases}
\SD(\sigma^1,R), & \text{if } R\neq \widehat R,\\
\SD(\sigma^2,R), & \text{if } R=\widehat R,
\end{cases}
\qquad
f^{\sigma^2}(R)=
\begin{cases}
\SD(\sigma^2,R), & \text{if } R\neq \widehat R,\\
\SD(\sigma^1,R), & \text{if } R=\widehat R.
\end{cases}
\]

Now define the convex lottery rule \(\widehat\varphi\in\Delta \mathcal{F}\) by
\[
\widehat\varphi_{f^{\sigma^1}}=\widehat\varphi_{f^{\sigma^2}}=\frac12,
\]
and
\[\text{for each }f\in \mathcal{F}\setminus\{f^{\sigma^1},f^{\sigma^2}\},\
\widehat\varphi_f=0.
\]
Note that the induced lottery rule \(\widehat\varphi:\mathcal R^N\to\Delta \M\) is such that for each $R\in\mathcal R^N$, $\widehat\varphi(R)=\frac12\,f^{\sigma^1}(R)+\frac12\,f^{\sigma^2}(R)$ (thus, it assigns the same probabilities as the RSD rule). Hence, the convex lottery rule $\widehat\varphi$ satisfies \textsl{probabilistic monotonicity}.

To see that \(\widehat\varphi\) violates \textsl{ex-post strategy-proofness}, it suffices to show that one deterministic rule in its support is not \textsl{strategy-proof}. We show that \(f^{\sigma^1}\) is not \textsl{strategy-proof}.

At profile \(\widehat R\), by definition,
\(
f^{\sigma^1}(\widehat R)=\SD(\sigma^2,\widehat R).
\)
Hence, agent~\(2\) chooses first and receives object \(a\), while agent \(1\) receives object
\(b\). Thus,
\(
f^{\sigma^1}(\widehat R)=(b,a).
\)

Now let agent~\(1\) report
\(
R_1':\ a \;P_1'\; c \;P_1'\; b,
\)
while agent~\(2\) keeps reporting \(\widehat R_2\). Since
\((R_1',\widehat R_2)\neq \widehat R\), the exceptional case no longer applies, and therefore
\(
f^{\sigma^1}(R_1',\widehat R_2)=\SD(\sigma^1,(R_1',\widehat R_2)).
\)
Thus, agent~\(1\) chooses first and receives object \(a\), so
\(
f^{\sigma^1}(R_1',\widehat R_2)=(a,b).
\)

Since under agent \(1\)'s true preferences, \(a \;\widehat{P}_1\; b\), we obtain
\(
f^{\sigma^1}(R_1',\widehat R_2)\; \widehat{P}_1\; f^{\sigma^1}(\widehat R).
\)
Hence, agent \(1\) can profitably manipulate \(f^{\sigma^1}\), implying that
\(f^{\sigma^1}\) is not \textsl{strategy-proof}.
Therefore, convex lottery rule \(\widehat\varphi\) violates \textsl{ex-post strategy-proofness}.\medskip

	Finally, Lemma~\ref{lemma:GSP} allows us to restate Theorem~\ref{th:main} for the subset of convex lottery rules in terms of \textsl{ex-post group strategy-proofness} instead of \textsl{probabilistic monotonicity}.
	
	\begin{corollary}[\textbf{Equivalence of \textsl{ex-post pairwise} and \textsl{ex-post Pareto efficiency} for convex lottery rules  on $\bm{\R^N}$}]\label{cor:main_convexLR} \ \\
		Let $N=\{1,\ldots,n\}$ and $O=\{o_1,\ldots,o_k\}$. Let $\varphi\in\Delta\mathcal{F}$ be a convex lottery rule  on $\R^N$ that satisfies \textsl{ex-post group strategy-proofness}.
		\begin{itemize}
			\item[\textbf{\emph{(a)}}] If $\sum_{o\in O} q_o > n$, then $\varphi$ is \textsl{ex-post pairwise efficient} and \textsl{ex-post non-wasteful} if and only if it is \textsl{ex-post Pareto efficient}.
			\item[\textbf{\emph{(b)}}] Otherwise, $\sum_{o\in O} q_o = n$, and $\varphi$ is \textsl{ex-post pairwise efficient} if and only if it is \textsl{ex-post Pareto efficient}.
		\end{itemize}
	\end{corollary}

\section{Ex-ante notions of non-manipulability}\label{ap:SD-notions}

In order to compare \textsl{probabilistic monotonicity} with various ex-ante non-manipulability concepts for probabilistic object allocation problems, we introduce some additional notation.\medskip

Given a profile $R\in\R^N$, a lottery rule $\varphi$, an agent $i\in N$, and an object $o\in O$, denote by $\varphi_{io}(R)$ the probability with which agent $i$ is assigned to object $o$ in lottery $\varphi(R)$, i.e., $$\varphi_{io}(R) = \sum_{\mu: \mu_i = o} \varphi_\mu(R).$$ Moreover, given an agent $i$, let $\varphi_i(R) = (\varphi_{io}(R))_{o\in O}$ denote agent $i$'s \textit{probabilistic allotment}, which equals the vector containing the assignment probabilities to all objects of agent $i$ at lottery $\varphi(R)$.\medskip

A common approach to extend agents' preferences over objects to preferences over probabilistic allotments is through \emph{stochastic dominance}  \citep[see][]{bogomolnaiamoulin2001}. In the following description we use the interpretation that an agent's preferences constitute a ranking of objects: rank 1 corresponds to the best object, rank 2 corresponds to the second best object, etc.
Given a lottery rule $\varphi$, an agent $i\in N$, and two profiles $R,R'\in \R^N$, probabilistic allotment $\varphi_i(R)$ \emph{stochastically dominates} probabilistic allotment $\varphi_i(R')$ if, for each rank $r$, the probability of being assigned to an object ranked $r$-th or better at $R_i$ is at least as large at $\varphi_i(R)$ as at $\varphi_i(R')$. Since agents have strict preferences, and each rank in the preferences contains exactly one object, $\varphi_i(R)$ stochastically dominates  $\varphi_i(R')$ if, for each object $o\in O$, $$\sum_{o':\,o'\mathbin{R_i} o} \varphi_{io'}(R) \geq \sum_{o':\,o'\mathbin{R_i} o} \varphi_{io'}(R').$$

Let $\varphi_i(R) \succeq^{sd}_i \varphi_i(R')$ denote that $\varphi_i(R)$ stochastically dominates $\varphi_i(R')$. Additionally, if $\varphi_i(R) \succeq^{sd}_i \varphi_i(R')$, and there exists an object $o\in O$ for which $\sum_{o':\,o'\mathbin{R_i} o} \varphi_{io'}(R) > \sum_{o':\,o'\mathbin{R_i} o} \varphi_{io'}(R')$, then $\varphi_i(R)$ \emph{strictly} stochastically dominates $\varphi_i(R')$, which we denote by $\varphi_i(R) \succ^{sd}_i \varphi_i(R')$. We use the prefix ``sd'' to shorten stochastic dominance.\medskip	

The following sd-notions of \textsl{strategy-proofness} \citep[see][]{bogomolnaiamoulin2001} model that no agent can benefit, in the stochastic dominance sense, from misreporting her preferences. Note that the first notion requires sd-comparability of an agent's allotment after her preference change, while the second notion does not require that.

\begin{definition}[\textbf{Sd-strategy-proofness}]\ \\
	A lottery rule $\varphi$ is \textsl{sd-strategy-proof} if for each $R\in \R^N$, each agent $i\in N$, and each preference relation $R'_i\in \R$, $\varphi_i(R) \succeq^{sd}_i \varphi_i(R'_i, R_{-i})$.
\end{definition}

\begin{definition}[\textbf{Weak sd-strategy-proofness}]\ \\
	A lottery rule $\varphi$ is \textsl{weakly sd-strategy-proof} if for each $R\in \R^N$, there does not exist an agent $i\in N$ and a preference relation $R'_i\in \R$ such that $\varphi_i(R'_i, R_{-i}) \succ^{sd}_i \varphi_i(R)$.
\end{definition}

\textsl{Sd-strategy-proofness} implies \textsl{weak sd-strategy-proofness}.\medskip

We consider two different extensions of \textsl{group strategy-proofness} from deterministic to lottery rules.
A first extension of \textsl{group strategy-proofness} was introduced by \cite{bade2016fairness}. Given a lottery rule, consider a scenario in which a group $S$ of agents misreport their preferences such that at least one of the agents in this group $S$ receives a different probabilistic allotment than under truth-telling. We say that a lottery rule is \textsl{sd-group strategy-proof} if, whenever this scenario occurs, there is at least one agent in this group $S$ who is strictly worse off, in the stochastic dominance sense, than under truth-telling.
	
	\begin{definition}[\textbf{Sd-group strategy-proofness}]\ \\
		A lottery rule $\varphi$ satisfies \emph{sd-group strategy-proofness} if for each profile $R\in \R^N$, and for each group of agents $S\subseteq N$ with preferences $R'_S \in \R^S$ for which there is an agent $i'\in S$ whose probabilistic allotment changes, i.e., $\varphi_{i'}(R) \neq \varphi_{i'}(R'_S, R_{-S})$, there exists at least one agent $i \in S$ such that $\varphi_i(R) \succ^{sd}_i \varphi_i(R'_S, R_{-S})$.
	\end{definition}

When restricting the set of deviating agents to singletons, \textsl{sd-group strategy-proofness} reduces to \textsl{sd-strategy-proofness}.\footnote{Indeed, when restricting the deviating set of agents to singletons, the conditions in the definition of \textsl{sd-group strategy-proofness} that the probabilistic allotment of the manipulating agent should differ after the manipulation and that truth-telling should \emph{strictly} sd-dominate manipulating jointly imply the condition in the definition of \textsl{sd-strategy-proofness} that truth-telling should \emph{weakly} sd-dominate manipulating.} \medskip
	
\cite{bade2016fairness} shows that no lottery rule satisfies \textsl{sd-group strategy-proofness}, \textsl{equal treatment of equals}, and \textsl{ex-post Pareto efficiency} (whereas \cite{basteck2024axiomatization} shows that RSD is characterized by these properties when replacing \textsl{sd-group strategy-proofness} by \textsl{probabilistic monotonicity}).\footnote{Note that the definitions of the \textsl{equal treatment of equals} properties in \cite{bade2016fairness} and \cite{basteck2024axiomatization} differ, but that the notion by \cite{basteck2024axiomatization} implies the notion by \cite{bade2016fairness} (and hence can still be used to obtain the impossibility result).}\medskip

Because \textsl{sd-group strategy-proofness} is a very demanding property, \cite{zhang2019efficient} proposes a less demanding extension of \textsl{group strategy-proofness} to lottery rules. A lottery rule is \textsl{weakly sd-group strategy-proof} if no group of agents can manipulate and all be strictly better off, in the stochastic dominance sense, than under truth-telling.

\begin{definition}[\textbf{Weak sd-group strategy-proofness}]\ \\
	A lottery rule $\varphi$ is \textsl{weakly sd-group strategy-proof} if for each $R\in \R^N$, there is no group of agents $S\subseteq N$ with preferences $R'_S \in \R^S$ such that for each $i\in S$, $\varphi_i(R'_S, R_{-S}) \succ^{sd}_i \varphi_i(R)$.
\end{definition}

When restricting the set of deviating agents to singletons, \textsl{weak sd-group strategy-proofness} reduces to \textsl{weak sd-strategy-proofness}. \medskip

\textsl{Sd-group strategy-proofness} implies \textsl{weak sd-group strategy-proofness}.\medskip

Whereas \textsl{monotonicity} and \textsl{group strategy-proofness} are equivalent for deterministic rules (Proposition~\ref{prop:equivalences}), we show that \textsl{probabilistic monotonicity} is logically independent of \textsl{(weak) sd-group strategy-proofness}.\medskip

On the one hand, the RSD rule is an example of a lottery rule that satisfies (\textsl{ex-post Pareto efficiency} and) \textsl{probabilistic monotonicity} \citep{basteck2024axiomatization}, but violates \textsl{(weak) sd-group strategy-proofness} \citep{bade2016fairness, zhang2019efficient}. To illustrate this, let $N=\{1,2,3\}, O = \{a,b,c\}$, and $q=(1,1,1)$. Consider the following two preference profiles, in which agents $1$ and $2$ manipulate by swapping objects $a$ and $b$ in their preferences. The fractions in parentheses indicate the assignment probability by the RSD rule at this profile:
\begin{center}
	\renewcommand{\arraystretch}{0.8}
	\begin{tabular}{cc}
		$R_1 = R_2$ & $R_3 $\\ \hline
		$a \;(\frac{1}{2})$ & $b \;(\frac{2}{3})$\\
		$b \;(\frac{1}{6})$ & $c \;(\frac{1}{3})$\\
		$c \;(\frac{1}{3})$ & $a \;(0)$			
\end{tabular} \qquad \qquad\begin{tabular}{cc}
$R'_1 = R'_2$ & $R_3 $\\ \hline
$b \;(\frac{1}{3})$ & $b \;(\frac{1}{3})$\\
$a \;(\frac{1}{2})$ & $c \;(\frac{2}{3})$\\
$c \;(\frac{1}{6})$ & $a \;(0)$		
\end{tabular} \end{center}

Agents 1 and 2 benefit from the manipulation from $(R_1, R_2)$ to $(R'_1, R'_2)$, as their probabilistic allotments at profile $(R'_1, R'_2, R_3)$ strictly stochastically dominate their probabilistic allotments at profile $R$. Hence, RSD violates \textsl{weak sd-group strategy-proofness}, and therefore also \textsl{sd-group strategy-proofness}.\medskip

On the other hand, the following lottery rule is an example of a rule that satisfies \textsl{sd-group strategy-proofness} (and therefore also \textsl{weak sd-group strategy-proofness}), but violates \textsl{probabilistic monotonicity}. Let $N=\{1,2,3\}, O=\{a,b,c\}$, and $q=(1,1,1)$. Let
$\widehat R=(\widehat R_1,\widehat R_2,\widehat R_3)\in\mathcal R^N$ be the preference profile such that,
for each $i\in N$, $\widehat R_i:\ a \mathbin{\widehat P_i} b \mathbin{\widehat P_i} c$. Consider the lottery rule $\varphi$ which outputs the following lottery:\vspace{-0.3cm}
\[\varphi(R) = \begin{cases}
	\frac{1}{3}(a,c,b) + \frac{1}{3}(b,a,c) +  \frac{1}{3}(c,b,a) &\text{ if } R = \widehat R,\\
	\frac{1}{3}(a,b,c) + \frac{1}{3}(b,c,a) + \frac{1}{3}(c,a,b) &\text{ if } R \neq \widehat R.
	\end{cases}\vspace{-0.3cm}\]
Clearly, every agent has a probability of $\frac{1}{3}$ of being assigned to every object, regardless of her submitted preferences. Hence, $\varphi$ satisfies \textsl{sd-group strategy-proofness}, and therefore also \textsl{weak sd-group strategy-proofness}. Nevertheless, for any profile $R\neq \widehat R$ such that $\widehat R$ is a monotonic transformation of $R$ at matching $(a,b,c)$, the weight of matching $(a,b,c)$ is lower at $\varphi(\widehat R)$ than at $\varphi(R)$, which violates \textsl{probabilistic monotonicity}.

	\section{Proof of Theorem~\ref{th:restricted}, Case~(a)}
	\label{app:proof_restricted}

	\begin{proof}[\textbf{Proof}]The ``if'' direction follows directly from \textsl{pairwise efficiency} and \textsl{non-wastefulness} being a relaxation of \textsl{Pareto efficiency}.\medskip
		
		To prove the ``only if'' direction, consider a deterministic rule $f$ on $\underline{\R}^N$ that satisfies \textsl{pairwise strategy-proofness}, \textsl{pairwise efficiency}, and \textsl{non-wastefulness}. Proposition~\ref{prop:equivalences} continues to hold on $\underline{\R}^N$ and hence, $f$ is \textsl{group strategy-proof} and \textsl{monotonic}. Suppose, by contradiction, that $f$ is not \textsl{Pareto efficient}. Hence, there exists a preference profile $R\in\underline{\R}^N$ and a matching $\mu\in\M$ such that $f(R)=\mu$, and $\mu$ is not \textsl{Pareto efficient}. Let $\nu\in\M$ denote a matching that Pareto dominates $\mu$.
		
		Recall that $f$ is defined on $\underline{\R}^N$ and $\sum_{o\in O} q_o > n$. Hence, $q_{\varnothing}>n-\sum_{o\in O\setminus\{\varnothing\}}q_o$ and all agents rank the null object last. Thus, by \textsl{non-wastefulness}, an agent only receives the null object $\varnothing$ when all ``real objects'' in $O\setminus\{\varnothing\}$ are allocated at $f(R)=\mu$. Thus,
		\begin{itemize}
			\item[]  an agent $i\in N$ who receives the null object at $\mu$ receives the null object at $\nu$, i.e., $\mu_i=\varnothing$ if and only if $\nu_i=\varnothing$;
			\item[] an agent $i\in N$ who receives a real object at $\mu$ receives a real object at $\nu$, i.e., $\mu_i\in O\setminus\{\varnothing\}$ if and only if $\nu_i\in O\setminus\{\varnothing\}$.
		\end{itemize}
		
		We partition the set of agents into three groups:
		\begin{itemize}
			\item[] $N_{\neq}$ is the subset of agents who receive a different real object in $O\setminus\{\varnothing\}$ at $\mu$ and $\nu$, i.e., $$N_{\neq} = \{i\in N: \mu_i \neq \nu_i \text{ and } \mu_i, \nu_i \neq \varnothing\},$$
			\item[] $N_{=}$ is the subset of agents who receive the same real object in $O\setminus\{\varnothing\}$ at $\mu$ and $\nu$, i.e., $$N_{=} = \{i\in N: \mu_i = \nu_i \neq \varnothing\},$$
			\item[] $N_\varnothing$ is the subset of agents who receive the null object at  $\mu$ and $\nu$, i.e., $$N_\varnothing = \{i\in N: \mu_i =\nu_i= \varnothing\}.$$
		\end{itemize}
		Note that if $N_\varnothing =\emptyset$, then the proof of Theorem~\ref{th:main} / Corollary~\ref{cor:det_rules} remains valid to also prove Theorem~\ref{th:restricted}. It is therefore without loss of generality to assume that $N_\varnothing \neq\emptyset$.\medskip
		
		Let $R'$ denote the preference profile in $\underline{\R}^N$ that is obtained from $R$ by
		\begin{itemize}
			\item moving for each agent $i\in N_{\neq}$, object $\nu_i$ to the top of her preferences and object $\mu_i$ to the second spot in her preferences, without changing the ranking of other objects, i.e.,
			\item[\textbf{-}] for each object $o\in O\setminus\{\nu_i\}$, $\nu_i \mathbin{P'_i} \mu_i \mathbin{R'_i} o$ and
			\item[\textbf{-}] for each pair of objects $o,o'\in O\setminus\{\nu_i,\mu_i\}$, $o\mathbin{R'_i} o'$ if and only if $o\mathbin{R_i}o'$;
			\item moving for each agent $i\in N_{=}$, object $\nu_i=\mu_i$ to the top of her preferences, without changing the ranking of other objects, i.e.,
			\item[\textbf{-}] for each object $o\in O$, $\nu_i = \mu_i \mathbin{R'_i} o$ and
			\item[\textbf{-}] for each pair of objects $o,o'\in O\setminus\{\nu_i=\mu_i\}$, $o\mathbin{R'_i} o'$ if and only if $o\mathbin{R_i}o'$;
			\item not changing the preferences of agents $i\in N_\varnothing$, i.e., $R'_i=R_i$.
		\end{itemize}
		
		Because $\nu$ Pareto dominates $\mu$, we have that for all agents $i \in N$, $\nu_i\mathbin{R_i}\mu_i$ and the new preference profile $R'$ is a monotonic transformation of $R$ at $\mu$. Hence, by \textsl{monotonicity}, $f(R') = f(R) = \mu$.\footnote{Note that, in contrast to the proof of Theorem~\ref{th:main} / Corollary~\ref{cor:det_rules}, matching $\nu$ is now not top-ranked by all agents at preference profile $R'$ (due to agents in $N_\varnothing$, who receive the null object, but would rather have a real object in $O\setminus\{\varnothing\}$).} \medskip
		
		Because $\nu$ Pareto dominates $\mu$, without loss of generality, there exists an improvement cycle involving agents $1,\ldots,\ell$ such that each agent~$j$ in the cycle is assigned to object $a_j = \mu_j$, but prefers object $a_{j-1} = \nu_j = \mu_{j-1}$ (modulo $\ell$). Without loss of generality, we also assume that the improvement cycle is a shortest improvement cycle, which implies that all improvement cycle objects $a_1,\ldots,a_\ell$ are mutually different real objects in $O\setminus\{\varnothing\}$, i.e., $|\{a_1,\ldots,a_\ell\}|=\ell$. Note that $\ell =2$ would mean that agents 1 and 2 would like to swap their allotments, contradicting \textsl{pairwise efficiency}. Hence, $\ell\geq 3$.\medskip
		
		We denote this improvement cycle by $C=\langle a_\ell, \ell, a_{\ell-1}, \ell-1, a_{\ell-2}, \ell-2, a_{\ell-3},\ldots, a_2, 2, a_1, 1, a_\ell\rangle$. It is without loss of generality to assume that $C$  is the only improvement cycle: if there were other improvement cycles, we could apply a monotonic transformation of preferences where all other improvement cycle agents would move their allotment at $\mu$ to the top of their preferences; matching $\mu$ would still be chosen at the monotonically transformed preference profile and $C$ would then be the unique improvement cycle. Thus, $$N_{\neq}=\{1,\ldots,\ell\},\ \ell \geq 3,$$ and implementing the improvement cycle at $\mu$ would result in $\nu$. Furthermore, assume that $$N_{=}=\{\ell+1,\ldots,\kappa\}$$ and $$N_{\varnothing}=\{\kappa +1,\ldots,n\}\neq\emptyset.$$

		Let $\succ: a_1 \succ a_2 \succ \ldots \succ a_{\ell} \succ a_{\ell + 1} \succ \ldots \succ a_{k}$ denote a common ranking of the objects in $O\setminus\{\varnothing\}$, which first ranks objects $a_1, \ldots, a_\ell$ belonging to improvement cycle $C$, and then arbitrarily ranks the remaining objects $a_{\ell+1}, \ldots, a_k$. In the remainder of the proof, for each agent $i\in N_{\neq}$, the labels $a_i = \mu_i$ and $a_{i-1} = \nu_i = \mu_{i-1}$ will be used interchangeably. \medskip
		
		Let $\hatR$ denote the preference profile in $\underline{\R}^N$ that is obtained from $R'$ by
		\begin{itemize}
			\item rearranging, for each agent~$i\in N\setminus N_\varnothing$, all objects that are ranked below object $\mu_i$ according to the common ranking~$\succ$, i.e.,
			\item[\textbf{-}] for each object $o\in O\setminus\{\nu_i, \mu_i, \varnothing\}$, $\nu_i \mathbin{\hatR_i} \mu_i \mathbin{\hatP_i}o\mathbin{\hatP_i}\varnothing$ and
			\item[\textbf{-}]  for each pair of objects $o,o'\in O\setminus\{\nu_i, \mu_i, \varnothing\}$, $o\mathbin{\hatP_i}o'$ if and only if $o \succ o'$;
			
			\item  rearranging the preferences of agents in $N_\varnothing$ such that they coincide with the common ranking, i.e.,
			\item[\textbf{-}] for each object $o\in O\setminus\{\varnothing\}$, $o \mathbin{\hatR_i} \varnothing$ and
			\item[\textbf{-}]for each pair of objects $o, o'\in O\setminus\{\varnothing\}$, $o\mathbin{\hatP_i}o'$ if and only if $o\mathbin{\succ}o'$.
		\end{itemize}
		
		Because for each agent $i\in N$, the position of $\mu_i$ remains the same under $R'$ and $\hatR$, and the null object is still ranked last for all agents, $\hatR\in \underline{\R}^N$ is a monotonic transformation of $R'$ at $\mu$. Hence, by \textsl{monotonicity}, $f(\hatR) = f(R') = \mu$. Figure~\ref{fig:tildeR} illustrates the resulting preference profile $\hatR$ for a problem with unit capacities, which implies $\kappa = k$.\medskip
		
		\begin{figure}[ht!]
			
			\centering
			
			\begin{tabular}{cccccccccccc}
				$\hatR_1$ & $\hatR_2$ & $\hatR_3$ & $\hatR_4$ & \ldots & $\hatR_\ell$ & $\hatR_{\ell+1}$ & \ldots & $\hatR_{k}$ & $\hatR_{k+1}$ & \ldots & $\hatR_{n}$\\ \hline
				$a_\ell$ & $ a_1$& $a_2$ & $a_3$ & \ldots & $a_{\ell-1}$ & $\underline{a_{\ell+1}}$ & \ldots & $\underline{a_{k}}$ & $a_1$ & \ldots & $a_1$\\
				$\underline{a_1}$ & $\underline{a_2}$ & $\underline{a_3}$ & $\underline{a_4}$ & \ldots & $\underline{a_\ell}$ & $a_1$ & \ldots & $a_1$ & $a_2$ & \ldots & $a_2$\\
				$a_2$ & $a_3$ & $a_1$ & $a_1$ &\ldots & $a_1$ & $a_2$ & \ldots & $a_2$ & $a_3$ & \ldots & $a_3$\\
				$a_3$ & $a_4$ & $a_4$ & $a_2$ & \ldots & $a_2$ & $a_3$ & \ldots & $a_3$ & $a_4$ & \ldots & $a_4$\\
				\vdots & \vdots & \vdots & \vdots & \vdots & \vdots & \vdots & \vdots & \vdots & \vdots & \vdots & \vdots \\
				$a_k$ & $a_k$ & $a_k$ & $a_k$ & \ldots & $a_k$ & $a_k$ & \ldots & $a_{k - 1}$ & $a_k$ & \ldots & $a_k$\\
				$\varnothing$ & $\varnothing$ & $\varnothing$ & $\varnothing$ & \ldots & $\varnothing$ & $\varnothing$ & \ldots & $\varnothing$ & $\underline{\varnothing}$ & \ldots & $\underline{\varnothing}$\\
				\multicolumn{6}{c}{\upbracefill}& \multicolumn{3}{c}{\upbracefill} & \multicolumn{3}{c}{\upbracefill}\\
				\multicolumn{6}{c}{$N_{\neq}$}& \multicolumn{3}{c}{$N_{=}$} & \multicolumn{3}{c}{$N_\varnothing$}\\
				\
			\end{tabular}
			\caption{Profile $\hatR$, where matching $\mu$ is underlined.\label{fig:tildeR}}
		\end{figure}	
		
		Denote the set of agents who rank real object $o\in O\setminus\{\varnothing\}$ first at $\hatR$ by $N_{o}$, i.e.,  $N_{o} = \{i\in N: \text{for all } o' \in O, o \mathbin{\hatR_i} o'\}$. We first observe that, for each object, the same number of copies is assigned at $\nu$ and at any \textsl{non-wasteful} matching $\mu'$ (including matching $\mu$).
		
		\begin{observation}
			\label{obs2}
			Let $\mu'$ be a \textsl{non-wasteful} matching. Then, for each object $o\in O\setminus\{\varnothing\}$,  $|\{i \in N: \mu'_i = o\}|=|\{i\in N:\nu_i = o\}|$, and, for each object $o\in O\setminus\{\varnothing, a_1\}$, this number equals $|N_o|$.
		\end{observation}	
		
		The proof of Observation~\ref{obs2} follows the same arguments as that of Observation~\ref{obs1} in the proof of Theorem~\ref{th:main}; the equality with $|N_o|$ requires $o\neq a_1$ because every agent in $N_\varnothing$ ranks $a_1$ first at $\hatR$, so that $|N_{a_1}|=|\{i\in N:\nu_i = a_1\}|+|N_\varnothing|$.\medskip
		
		Let $\tilR$ denote the preference profile in $\underline{\R}^N$ that is obtained from $\hatR$ by
		\begin{itemize}
			\item  rearranging, for each agent~$i \in N_{\neq}=\{1,\ldots,\ell\}$, all objects that are ranked below object $\nu_i$ (including $\mu_i$) according to the common ranking~$\succ$, i.e.,
			\item[\textbf{-}] for each object $o\in O\setminus\{\nu_i,\varnothing\}$, $\nu_i \mathbin{\tilP_i} o\mathbin{\tilP_i} \varnothing$ and
			\item[\textbf{-}] for each pair of objects $o,o'\in O\setminus\{\nu_i, \varnothing\}$, $o\mathbin{\tilP_i}o'$ if and only if $o \succ o'$;
			
			\item  not changing the preferences of agents $i\in N\setminus N_{\neq}=\{\ell+1\ldots,n\}$, i.e.,  $\tilR_i=\hatR_i$.
		\end{itemize}
		
		In order to obtain the desired contradiction, we will stepwise transform preferences $\hatR_1,\ldots,\hatR_\ell$ into preferences $\tilR_1,\ldots,\tilR_\ell$,\footnote{Note that this is a rather different proof step, compared to the proof of Theorem~\ref{th:main}  / Corollary~\ref{cor:det_rules}, due to working on the restricted domain $\underline{\R}^N$ instead of the general domain $\R^N$.} and show that throughout this stepwise procedure, the matching chosen by $f$ equals $\mu$. After $\ell$ steps, we then will have shown that $f(\tilR) = f(\hatR) = \mu$ and obtain the desired contradiction. \medskip

		\noindent\textbf{Steps~1 and 2.} Because of our definition of the agents' labels and of the common ranking~$\succ$, at agent~1's preferences $\hatR_1$, all objects, except $\nu_1 = a_\ell$  are already ranked according to the common ranking $\succ$, i.e., $\hatR_1 = \tilR_1$.
		Similarly,  at agent 2's preferences $\hatR_2$, all objects are already ranked according to the common ranking $\succ$ (including $\nu_2 = a_1$), i.e., $\hatR_2 = \tilR_2$. Recall that the same is true for all agents $i \in N_{=}\cup N_{\varnothing}$, i.e., $\hatR_i = \tilR_i$.
		
		The following fact will be useful in Step~3: since $a_1\succ a_2$, by the definition of $\hatR$, the only agents that prefer $a_2$ to $a_1$ at $\hatR$ are agents in $N_{a_2}$.
		We now proceed to Step~3.\medskip
		
		\noindent \textbf{Step~3.} Note that at $\hatR_3$, $a_3=\mu_3$ and $a_2\mathbin{\hatP_3}a_3\mathbin{\hatP_3}a_1$, while $a_1 \succ a_3$. Let $\hatR'_3\in \underline{\R}$ be preferences that are obtained from $\hatR_3$ by
		\begin{itemize}
			\item swapping the positions of adjacent objects $a_1$ and $a_3$, i.e.,
			\item[\textbf{-}] for each object $o\in O\setminus\{a_1,a_2,a_3,\varnothing\}$, $a_2\mathbin{\hatP'_3}a_1\mathbin{\hatP'_3}a_3\mathbin{\hatP'_3}o\mathbin{\hatP'_3}\varnothing$ and
			\item[\textbf{-}] for each pair of objects $o,o'\in O\setminus\{a_1,a_2,a_3, \varnothing\}$, $o\mathbin{\hatP'_3}o'$ if and only if $o \succ o'$.
		\end{itemize}
		Let $\hatR' = (\hatR'_3, \hatR_{-3})$ and $\mu'=f(\hatR')$.\medskip
		
		At $\hatR'_3$, all objects, except $\nu_3 = a_2$  are now ranked according to the common ranking $\succ$, i.e., $\hatR'_3=\tilR_3$. Recall that the same is true for each agent $i\in \{1,2\}\cup N_{=}\cup N_{\varnothing}$, i.e., $\hatR'_i=\tilR_i$. \medskip
		
		If $\ell > 3$, then the following fact will be useful in Step~4: since $a_1\succ a_2\succ a_3$, by the definition of $\hatR'$, the only agents that prefer object  $a_3$ to objects in $\{a_1,a_2\}$ at $\hatR'$ are agents in $N_{a_3}$.\medskip
		
		Agent~$3$ cannot receive object $\nu_3 = a_2$ at $\mu'$, because the change from $\hatR_3$ to $\hatR'_3$ would then violate \textsl{strategy-proofness}. Similarly, agent~$3$ cannot receive any object she ranks strictly lower than $\mu_3 = a_3$ at $\hatR'_3$, because the change from $\hatR'_3$ to $\hatR_3$ would then violate \textsl{strategy-proofness}. Hence, $\mu'_3\in\{a_1,a_3\}$.\medskip
		
		Assume that  at $\mu'$, agent~$3$ receives object $a_1$. Then, by Observation~\ref{obs2}, there must be an agent $j\not\in N_{a_2}$ such that $\mu'_j=a_2=\nu_3$.  Since the only agents that prefer $a_2$ to $a_1$ are agents in $N_{a_2}$, $a_1\mathbin{\hatP'_j}a_2=\mu'_j$. Thus, agents 3 and $j$ would like to swap their allotments, contradicting \textsl{pairwise efficiency}.\medskip
		
		Hence, at $\mu'$, agent~3 still receives object $a_3 = \mu_3$. As the allotment of agent~3 is the same for preference profiles $\hatR$ and $\hatR'$, by \textsl{non-bossiness}, $f$ must still select matching $\mu$ at preference profile $\hatR'$, such that $\mu'=f(\hatR') = f(\hatR) = \mu$. To simplify notation, redefine preference profile $\hatR := \hatR'$ and consider $f(\hatR)=\mu$.  If $\ell = 3$, then $\hatR=\tilR$ and $f(\hatR)= f(\tilR)= \mu$. Otherwise, we proceed to Step~4.\medskip
		
		Recall that at preferences $\hatR_1,\hatR_2, \hatR_3$, all objects (except respectively $a_\ell, a_1, a_2$)  are already ranked according to the common ranking $\succ$. Recall that the same is true for all agents $i \in N_{=}\cup N_{\varnothing}$, i.e., $\hatR_i = \tilR_i$.\medskip

		\noindent \textbf{Step~4.} Note that at $\hatR_4$, $a_4=\mu_4$ and $a_3\mathbin{\hatP_4}a_4\mathbin{\hatP_4}a_1\mathbin{\hatP_4}a_2$, while $a_1 \succ a_2 \succ a_4$. Let $\hatR'_4\in \underline{\R}$ be preferences that are obtained from $\hatR_4$ by
		\begin{itemize}
			\item rearranging objects $a_1,a_2,a_4$  according to the common ranking~$\succ$, i.e.,
			\item[\textbf{-}] for each object $o\in O\setminus\{a_1,a_2,a_3,a_4,\varnothing\}$, $a_3\mathbin{\hatP'_4}a_1\mathbin{\hatP'_4}a_2\mathbin{\hatP'_4}a_4\mathbin{\hatP'_4}o \mathbin{\hatP'_4}\varnothing$ and
			\item[\textbf{-}] for each pair of objects $o,o'\in O\setminus\{a_1,a_2,a_3,a_4,\varnothing\}$, $o\mathbin{\hatP'_4}o'$ if and only if $o \succ o'$.
		\end{itemize}
		Let $\hatR' = (\hatR'_4, \hatR_{-4})$ and $\mu'=f(\hatR')$. Figure~\ref{fig:hatR_4} illustrates the resulting preference profile $\hatR'$ for a problem with unit capacities, which implies $\kappa = k$.\medskip
		
		\begin{figure}[ht!]
			
			\centering
			
			\begin{tabular}{ccccccccccccc}
				$\tilR_1$ & $\tilR_2$ & $\tilR_3$ & $\hatR'_4$ & $\hatR_5$ & \ldots & $\hatR_\ell$ & $\tilR_{\ell+1}$ & \ldots & $\tilR_{k}$ & $\tilR_{k+1}$ & \ldots & $\tilR_{n}$\\ \hline
				$a_\ell$ & $ a_1$& $a_2$ & $a_3$ & $a_4$ & \ldots & $a_{\ell-1}$ & $\underline{a_{\ell+1}}$ & \ldots & $\underline{a_{k}}$ & $a_1$ & \ldots & $a_1$\\
				$\underline{a_1}$ & $\underline{a_2}$ & $a_1$ & $a_1$ & $\underline{a_5}$ & \ldots & $\underline{a_\ell}$ & $a_1$ & \ldots & $a_1$ & $a_2$ & \ldots & $a_2$\\
				$a_2$ & $a_3$ & $\underline{a_3}$ & $a_2$ & $a_1$ & \ldots & $a_1$ & $a_2$ & \ldots & $a_2$ & $a_3$ & \ldots & $a_3$\\
				$a_3$ & $a_4$ & $a_4$ & $\underline{a_4}$ & $a_2$ & \ldots & $a_2$ & $a_3$ & \ldots & $a_3$ & $a_4$ & \ldots & $a_4$\\
				\vdots & \vdots & \vdots & \vdots & \vdots & \vdots & \vdots & \vdots & \vdots & \vdots & \vdots & \vdots & \vdots \\
				$a_k$ & $a_k$ & $a_k$ & $a_k$ & $a_k$ & \ldots & $a_k$ & $a_k$ & \ldots & $a_{k - 1}$ & $a_k$ & \ldots & $a_k$\\
				$\varnothing$ & $\varnothing$ & $\varnothing$ & $\varnothing$ & $\varnothing$ & \ldots & $\varnothing$ & $\varnothing$ & \ldots & $\varnothing$ & $\underline{\varnothing}$ & \ldots & $\underline{\varnothing}$\\
				\multicolumn{7}{c}{\upbracefill}& \multicolumn{3}{c}{\upbracefill} & \multicolumn{3}{c}{\upbracefill}\\
				\multicolumn{7}{c}{$N_{\neq}$}& \multicolumn{3}{c}{$N_{=}$} & \multicolumn{3}{c}{$N_\varnothing$}\\
				\
			\end{tabular}
			\caption{Step~4 preference profile $\hatR'$, where matching $\mu$ is underlined. \label{fig:hatR_4}}
		\end{figure}

		At $\hatR'_4$, all objects, except $\nu_4 = a_3$  are now ranked according to the common ranking $\succ$, i.e., $\hatR'_4=\tilR_4$. Recall that the same is true for all $i\in \{1,2,3\}\cup N_{=}\cup N_{\varnothing}$, i.e., $\hatR'_i=\tilR_i$. \medskip
		
		If $\ell > 4$, then the following fact will be useful in Step~5: since $a_1\succ a_2\succ a_3\succ a_4$, by the definition of $\hatR'$, the only agents that prefer object  $a_4$ to objects in $\{a_1,a_2,a_3\}$ are agents in $N_{a_4}$.\medskip
		
		Agent~$4$ cannot receive object $\nu_4 = a_3$ at $\mu'$, because the change from $\hatR_4$ to $\hatR'_4$ would then violate \textsl{strategy-proofness}. Similarly, agent~$4$ cannot receive any object she ranks strictly lower than $\mu_4 = a_4$ at $\hatR'_4$, because the change from $\hatR'_4$ to $\hatR_4$ would then violate \textsl{strategy-proofness}.  Hence, $\mu'_4\in\{a_1,a_2,a_4\}$.\medskip
		
		Assume that  at $\mu'$, agent~4 receives an object in $\{a_1,a_2\}$. Then, by Observation~\ref{obs2}, there must be an agent $j\not\in N_{a_3}$ such that $\mu'_j=a_3=\nu_4$. Since the only agents that prefer object  $a_3$ to objects in $\{a_1,a_2\}$ are agents in $N_{a_3}$, $a_1,a_2\mathbin{\hatP'_j}a_3=\mu'_j$. Thus, agents 4 and $j$ would like to swap their allotments, contradicting \textsl{pairwise efficiency}.\medskip
		
		Hence, at $\mu'$, agent~4 still receives object $a_4 = \mu_4$. As the allotment of agent~4 is the same for preference profiles $\hatR$ and $\hatR'$, by \textsl{non-bossiness}, $f$ must still select matching $\mu$ at preference profile $\hatR'$, such that $\mu' = f(\hatR') = f(\hatR) = \mu$. To simplify notation, redefine preference profile $\hatR := \hatR'$ and consider $f(\hatR)=\mu$.  If $\ell = 4$, then $\hatR=\tilR$ and $f(\hatR)= f(\tilR)= \mu$. Otherwise, we proceed to Step~5. \medskip 		
		
		Steps after Step~4 take the following form.\medskip		
		
		If $\ell \geq s$, after Step~$s-1$ the following fact is true: since $a_1\succ \ldots \succ a_{s-1}$, by the definition of $\hatR'$, the only agents that prefer object  $a_{s-1}$ to objects in $\{a_1,\ldots,a_{s-2}\}$ are agents in $N_{a_{s-1}}$. Furthermore, for each agent  $i\in \{1,\ldots, s-1\}\cup N_{=}\cup N_{\varnothing}$, $\hatR'_i=\tilR_i$. \medskip

		\noindent \textbf{Step~$\bm{s}$.} Note that at $\hatR_s$, $a_s=\mu_s$ and $a_{s-1}\mathbin{\hatP_s}a_s\mathbin{\hatP_s}a_1\mathbin{\hatP_s}\ldots \mathbin{\hatP_s}a_{s-2}$, while $a_1 \succ a_2 \succ \ldots \succ a_{s-1}\succ a_s$. Let $\hatR'_s\in \underline{\R}$ be preferences that are obtained from $\hatR_s$ by
		\begin{itemize}
			\item rearranging objects $a_1,a_2,\ldots,a_{s-2},a_s$  according to the common ranking~$\succ$, i.e.,
			\item[\textbf{-}] for each object $o\in O\setminus\{a_1,a_2,\ldots,a_{s-2},a_{s-1},a_s,\varnothing\}$, $a_{s-1}\mathbin{\hatP'_s}a_1\mathbin{\hatP'_s}a_2\mathbin{\hatP'_s}\ldots\mathbin{\hatP'_s}a_{s-2}\mathbin{\hatP'_s}a_s\mathbin{\hatP'_s}o\mathbin{\hatP'_s}\varnothing$ and
			\item[\textbf{-}] for each pair of objects $o,o'\in O\setminus\{a_1,a_2,\ldots,a_{s-2},a_{s-1},a_s,\varnothing\}$, $o\mathbin{\hatP'_s}o'$ if and only if $o \succ o'$.
		\end{itemize}
		Let $\hatR' = (\hatR'_s, \hatR_{-s})$ and $\mu'=f(\hatR')$.\medskip
		
		At $\hatR'_s$, all objects, except $\nu_s = a_{s-1}$  are now ranked according to the common ranking $\succ$, i.e., $\hatR'_s=\tilR_s$. Recall that the same is true for all $i\in \{1,\ldots,s-1\}\cup N_{=}\cup N_{\varnothing}$, i.e., $\hatR'_i=\tilR_i$. \medskip
		
		If $\ell > s$, then the following fact will be useful in Step~$s+1$: since $a_1\succ a_2\succ \ldots\succ a_{s}$, by the definition of $\hatR'$, the only agents that prefer object  $a_{s}$ to objects in $\{a_1,\ldots,a_{s-1}\}$ are agents in $N_{a_{s}}$.\medskip
		
		Agent~$s$ cannot receive object $\nu_s = a_{s-1}$ at $\mu'$, because the change from $\hatR_s$ to $\hatR'_s$ would then violate \textsl{strategy-proofness}. Similarly, agent~$s$ cannot receive any object she ranks strictly lower than $\mu_s = a_s$ at $\hatR'_s$, because the change from $\hatR'_s$ to $\hatR_s$ would then violate \textsl{strategy-proofness}.  Hence, $\mu'_s\in\{a_1,a_2,\ldots,a_{s-2},a_s\}$.\medskip
		
		Assume that  at $\mu'$, agent~$s$ receives an object in $\{a_1,a_2,\ldots,a_{s-2}\}$. Then, by Observation~\ref{obs2}, there must be an agent $j\not\in N_{a_{s-1}}$ such that $\mu'_j=a_{s-1}=\nu_s$. Since the only agents that prefer object  $a_{s-1}$ to objects in $\{a_1,a_2,\ldots,a_{s-2}\}$ are agents in $N_{a_{s-1}}$, $a_1,a_2,\ldots,a_{s-2} \mathbin{\hatP'_j} a_{s-1}=\mu'_j$. Thus, agents $s$ and $j$ would like to swap their allotments, contradicting \textsl{pairwise efficiency}.\medskip
		
		Hence, at $\mu'$, agent~$s$ still receives object $a_s = \mu_s$. As the allotment of agent~$s$ is the same for preference profiles $\hatR$ and $\hatR'$, by \textsl{non-bossiness}, $f$ must still select matching $\mu$ at preference profile $\hatR'$, such that $\mu'= f(\hatR') = f(\hatR) = \mu$. To simplify notation, redefine preference profile $\hatR := \hatR'$  and consider $f(\hatR)=\mu$.  If $\ell = s$, then $\hatR=\tilR$ and $f(\hatR)= f(\tilR)= \mu$. Otherwise, we proceed to Step~$s+1$.\medskip	
		
		Finally, Step~$\ell$ ends with $\hatR=\tilR$ and $f(\hatR)= f(\tilR)= \mu$. Figure~\ref{fig:R_succ} illustrates the resulting preference profile $\tilR$ for a problem with unit capacities, which implies $\kappa = k$.\medskip
		
		\begin{figure}[ht!]
			
			\centering
			
			\begin{tabular}{cccccccccccc}
				$\tilR_1$ & $\tilR_2$ & $\tilR_3$ & $\tilR_4$ & \ldots & $\tilR_\ell$ & $\tilR_{\ell+1}$ & \ldots & $\tilR_{k}$ & $\tilR_{k+1}$ & \ldots & $\tilR_{n}$\\ \hline
				$a_\ell$ & $ a_1$& $a_2$ & $a_3$ & \ldots & $a_{\ell-1}$ & $\underline{a_{\ell+1}}$ & \ldots & $\underline{a_{k}}$ & $a_1$ & \ldots & $a_1$\\
				$\underline{a_1}$ & $\underline{a_2}$ & $a_1$ & $a_1$ & \ldots & $a_1$ & $a_1$ & \ldots & $a_1$ & $a_2$ & \ldots & $a_2$\\
				$a_2$ & $a_3$ & $\underline{a_3}$ & $a_2$ & \ldots & $a_2$ & $a_2$ & \ldots & $a_2$ & $a_3$ & \ldots & $a_3$\\
				$a_3$ & $a_4$ & $a_4$ & $\underline{a_4}$ & \ldots & $a_3$ & $a_3$ & \ldots & $a_3$ & $a_4$ & \ldots & $a_4$\\				
				\vdots & \vdots & \vdots & \vdots & \vdots & \vdots & \vdots & \vdots & \vdots & \vdots & \vdots & \vdots \\
				
				$a_{\ell-1}$ & $a_{\ell}$ & $a_{\ell}$ & $a_{\ell}$ & \ldots & $\underline{a_{\ell}}$ & $a_{\ell-1}$ & \ldots & $a_{\ell-1}$ & $a_{\ell-1}$ & \ldots & $a_{\ell-1}$\\
				$a_{\ell+1}$ & $a_{\ell+1}$ & $a_{\ell+1}$ & $a_{\ell+1}$ & \ldots & $a_{\ell+1}$ & $a_{\ell}$ & \ldots & $a_{\ell}$ & $a_{\ell}$ & \ldots & $a_{\ell}$\\
				
				\vdots & \vdots & \vdots & \vdots& \vdots & \vdots & \vdots & \vdots & \vdots & \vdots & \vdots & \vdots \\
				
				$a_k$ & $a_k$ & $a_k$ & $a_k$ & \ldots & $a_k$ & $a_k$ & \ldots & $a_{k - 1}$ & $a_k$ & \ldots & $a_k$\\
				$\varnothing$ & $\varnothing$ & $\varnothing$ & $\varnothing$ & \ldots & $\varnothing$ & $\varnothing$ & \ldots & $\varnothing$ & $\underline{\varnothing}$ & \ldots & $\underline{\varnothing}$\\
				\multicolumn{6}{c}{\upbracefill}& \multicolumn{3}{c}{\upbracefill} & \multicolumn{3}{c}{\upbracefill}\\
				\multicolumn{6}{c}{$N_{\neq}$}& \multicolumn{3}{c}{$N_{=}$} & \multicolumn{3}{c}{$N_\varnothing$}\\
				\
			\end{tabular}
			\caption{Profile $\tilR$ after Step~$\ell$, where matching $\mu$ is underlined. \label{fig:R_succ}}
		\end{figure}	
		
		Note that at $\mu$,  $a_\ell\mathbin{\tilP_1}a_1=\mu_1$ and $a_1\mathbin{\tilP_\ell}a_\ell=\mu_\ell$. Thus, agents 1 and $\ell$ would like to swap their allotments, contradicting \textsl{pairwise efficiency}.			
	\end{proof}

\end{document}